\documentclass[12pt]{aastex61}
\usepackage{amsmath}

\accepted{ for publication in the ApJS, 2023-05-31}
\submitjournal{ApJS}
\shorttitle{Spectral Line Survey of Ionized Jet Candidates}
\shortauthors{Sanchez-Tovar, et al.}

\newcommand {\kms }{$\,$km s$^{-1}$}
\newcommand {\JyB }{$\,$Jy beam$^{-1}$}

\usepackage{graphicx}

\begin{document}

\correspondingauthor{E. D. Araya}
\email{ed-araya@wiu.edu}

\title{Broadband VLA Spectral Line Survey of a Sample of Ionized Jet Candidates}

\author{E. Sanchez-Tovar}
\affiliation{Physics Department, Western Illinois University,
1 University Circle, Macomb, IL 61455, USA.}

\author{E. D. Araya}
\affiliation{Physics Department, Western Illinois University,
1 University Circle, Macomb, IL 61455, USA.}
\affiliation{New Mexico Institute of Mining and Technology, Physics Department, 801 Leroy Place, Socorro, NM 87801, USA.}

\author{V. Rosero}
\affiliation{National Radio Astronomy Observatory, 1003 Lopezville Road, Socorro, NM 87801, USA.}

\author{P. Hofner}
\altaffiliation{P. Hofner is also an Adjunct Astronomer at the National Radio Astronomy Observatory,
1003 Lopezville Road, Socorro, NM 87801, USA.}
\affiliation{New Mexico Institute of Mining and Technology, Physics Department, 801 Leroy Place, Socorro, NM 87801, USA.}

\author{S. Kurtz}
\affiliation{Instituto de Radioastronom\'{i}a y Astrof\'{i}sica,
Universidad Nacional Aut\'{o}noma de M\'{e}xico, Antig. Carr. a P\'{a}tzcuaro 8701, 58089, Morelia, Michoac\'{a}n, Mexico.}

\begin{abstract}

The study of the interaction between ionized jets, molecular outflows and their environments is critical to understanding high-mass star formation, especially because jets and outflows are thought to be key in the transfer of angular momentum outwards from accretion disks. We report a low-spectral resolution VLA survey for hydrogen radio recombination lines, OH, NH$_3$, and CH$_3$OH lines toward a sample of 58 high-mass star forming regions that contain numerous ionized jet candidates. The observations are from a survey designed to detect radio continuum; the novel aspect of this work is to search for spectral lines in broadband VLA data (we provide the script developed in this work to facilitate exploration of other datasets). We report detection of 25$\,$GHz CH$_3$OH transitions toward ten sources; five of them also show NH$_3$ emission. We found that most of the sources detected in CH$_3$OH and NH$_3$ have been classified as ionized jets or jet candidates and that the emission lines are coincident with, or very near ($\lesssim 0.1\,$pc) these sources, hence, these molecular lines could be used as probes of the environment near the launching site of jets/outflows. No radio recombination lines were detected, but we found that the RMS noise of stacked spectra decreases following the radiometer equation. Therefore, detecting radio recombination lines in a sample of brighter free-free continuum sources should be possible. This work demonstrates the potential of broadband VLA continuum observations as low-resolution spectral line scans.

\end{abstract}

\keywords{ISM: jets and outflows $-$ stars: formation $-$ ISM: molecules and recombination lines $-$ radio lines: ISM $-$ techniques: interferometric}

\section{INTRODUCTION}\label{sec:intro}

High-mass stars (M $\ga$ 8$\,$M$_{\odot}$) have widespread effects in the medium where they form and are likely a key factor that regulates further star formation in their natal clouds by feedback mechanisms, from the development of H{\small II} regions to supernovae (e.g., \citealt{Zinnecker_2007ARA&A..45..481Z}). During the earliest phases of formation, high-mass protostars can drive large ($\gtrapprox 1\,$pc) and massive ($\gtrapprox 10\,$M$_{\odot}$) molecular outflows that inject mechanical energy into the natal environment (e.g., \citealt{Rodriguez_2021RNAAS...5...70R}, \citealt{Torii_2017ApJ...840..111T}, \citealt{Bally_2016ARA&A..54..491B}, \citealt{Arce_2007prpl.conf..245A}). At smaller scales ($\sim 10^4\,$a.u.), high-sensitivity radio continuum observations find evidence that young high-mass protostars also drive ionized jets (e.g., \citealt{Purser_2021MNRAS.504..338P}, \citealt{Purser_2016MNRAS.460.1039P}, \citealt{Rosero_2019ApJ...880...99R}). In many cases, ionized jets are found to be approximately co-linear with large-scale molecular outflows (e.g., \citealt{Araya_2007ApJ...669.1050A}). Therefore, a connection between molecular outflows and ionized jets is expected, including the possibility that ionized jets can drive outflows. This idea is supported by observations that suggest a correlation between the momentum rates of molecular outflows and partially optically thick ionized jets (e.g., \citealt{Rosero_2019ApJ...880...99R}).

Spectral line studies are needed to detect and characterize molecular gas to explore the connection between ionized jets and the origin of outflows, particularly at the interphase of ionized jets. Traditionally, studies of radio continuum and spectral lines involve different observing settings (or independent runs), one to measure radio continuum and one to observe specific spectral lines. However, with the WIDAR correlator of the NSF's Karl G. Jansky Very Large Array (VLA), continuum mode observations are effectively {\it simultaneous} spectral line scans, albeit with broad channel widths. Spectral line detections from continuum mode observations have the additional advantage of having the nearly-identical baseline coverage, i.e., distribution of complex visibilities in the $(u,v)$-plane. Therefore, images of radio continuum emission (including ionized jets) can be directly compared to spectral line detections at similar angular resolution.

In this work we explore if narrow and bright maser lines or broad and weaker thermal lines\footnote{We refer to {\it thermal} lines those that originate from gas in local thermodynamic equilibrium (LTE) conditions as well as from gas close to LTE (often referred to as {\it quasi-thermal} lines in the literature, e.g., \citealt{Araya_2008ApJ...675..420A}).} from hydroxyl (OH), ammonia (NH$_3$) and methanol (CH$_3$OH) are detectable in low spectral resolution surveys obtained as a byproduct of continuum mode observations, as transitions from these species have been detected in high-mass star forming regions (e.g., \citealt{Tan_2020MNRAS.497.1348T}, \citealt{Henkel_2013A&A...549A..90H}). Transitions of CH$_3$OH and NH$_3$ are particularly interesting as masers have been detected together with thermal components of the same transitions, which is uncommon in other widespread maser species such as H$_2$O (e.g., \citealt{Towner_2017ApJS..230...22T}, \citealt{Goddi_2015A&A...573A.109G}, \citealt{Zhang_Ho_1995ApJ...450L..63Z}). Low spectral resolution surveys could also result in the detection of radio recombination lines (RRLs), particularly using stacking techniques (e.g., \citealt{Beuther_2016A&A...595A..32B}, \citealt{Liu_2013AJ....146...80L}).

We evaluate the idea of VLA continuum-mode observations as broadband spectral line scans by exploring the continuum observations of high-mass star-forming regions conducted by our group with the VLA \citep{Rosero_2016ApJS..227...25R}. In \cite{Rosero_2019ApJ...880...99R}, we characterized the sample of radio continuum sources detected in our main survey, in which we found a high fraction of ionized jet sources (including candidates). This paper explores whether transitions of NH$_3$, CH$_3$OH, and OH, as well as RRLs can be detected as by-products of the continuum observations. In Section~\ref{sec:obs} we describe the data used in this project as well as the algorithm developed to search for spectral lines from specific species and to stack spectral windows (SPWs) at RRL frequencies. In Section~\ref{sec:results} we discuss our detections of NH$_3$ and CH$_3$OH lines, as well as the non-detections of excited OH and RRLs. A summary is presented in Section~\ref{sec:summary}. The novel idea presented in this paper, i.e., the use of observations designed for radio-continuum measurements as low resolution spectral line scans, is applicable to any VLA broadband dataset. We provide a general purpose script that can be used to search for spectral lines from other VLA observations.
 
\newpage

\section{DATASET and SCRIPT DEVELOPMENT}\label{sec:obs}

Between 2010 and 2014, our group conducted an extensive survey for weak radio continuum sources toward 58 high-mass star-forming regions with the Karl G. Jansky Very Large Array (VLA) \citep{Rosero_2016ApJS..227...25R}. The sources in \cite{Rosero_2016ApJS..227...25R} were selected based on no previous radio continuum detection (or very weak detection at the $\sim 1\,$mJy level), i.e., a sample consistent with early evolutionary stages (see \citealt{Rosero_2016ApJS..227...25R} and \citealt{Rosero_2019ApJ...880...99R} for a detailed discussion of the sample). The main goal of the project was to investigate the nature of the weak continuum sources, which resulted in the detection of 70 sources associated with 1.2$\,$mm dust clumps (\citealt{Beuther_2002ApJ...566..945B}, \citealt{Rathborne_2006ApJ...641..389R}; see details in \citealt{Rosero_2016ApJS..227...25R}). We found that a significant fraction of them ($\sim 30\%$) may be tracing ionized jets \citep{Rosero_2019ApJ...880...99R}. The observations were conducted at C and K bands, using scaled arrays to achieve similar angular resolutions, i.e., C-band observations were conducted in A configuration and the K-band observations were obtained in B configuration. In each band, two sets of 8 spectral windows (128$\,$MHz bandwidth in most cases) were simultaneously observed, each set covered a bandwidth of 1024$\,$MHz centered at 4.9$\,$GHz (C-band), 7.4$\,$GHz (C-band), 20.9$\,$GHz (K-band) and 25.5$\,$GHz (K-band). Each spectral window comprised 64 channels. During the several years of the project, some variations were adopted in the spectral setups, and thus, not all frequencies were always observed in the same broadband continuum setup. Further details of the observation setup, including the results of the continuum observations, are reported in \cite{Rosero_2016ApJS..227...25R}.

Even though we are interested in investigating spectral lines in our specific continuum dataset, our approach is applicable to any pipeline calibrated visibility file (also known as Measurement Set, MS) from the VLA archive. Hence, we implemented the method of searching and imaging spectral windows with potentially interesting spectral lines in a {\sc Python} script accessible through GitHub\footnote{\url{https://github.com/AstroLab-WIU/Spectral_Line_Search_and_Stack}} (see installation and usage details in Appendix~\ref{sec:appendix}). The script uses tasks and functions from the NRAO software package {\sc CASA}\footnote{The code was tested in {\sc CASA} version 5.1, 5.6 and 6.2.}. Given that stacking of spectral lines may be needed for specific applications, e.g., search for RRLs, our script has two main components: 1) identification/imaging of individual spectral windows, and 2) stacking of spectral line cubes. Specifically:

\noindent - Step 1 (identification/imaging): The script generates a report of metadata of the MS file, including the frequency range associated with each spectral window. The script then compares the frequency ranges with a file of transitions of interest and runs the task {\tt tclean} to create cubes of the spectral windows in the LSRK reference frame using the rest frequencies. Given the large channel width of the observations, spectral line shifts due to Earth's rotation during an observing run are a small fraction of the channel width, and therefore, do not significantly contribute to spectral line dilution.\footnote{See discussion of velocity reference frames and Doppler correction in \newline \url{https://science.nrao.edu/facilities/vla/docs/manuals/obsguide/modes/line}.}

\noindent - Step 2 (stacking): Continuum subtraction is done in the image plane by fitting a zero order polynomial baseline to a subset of channels of each spectral cube. The script regrids each cube to have the effective channel width (in velocity) of the lowest frequency spectral window to be stacked, i.e., regrids to the lowest velocity resolution. Finally, the script generates the stacked cube using the RMS to weight the relative contribution of each cube (further details are included in Appendix~\ref{sec:appendix}).

\begin{figure}[ht]
  \begin{minipage}[c]{0.5\linewidth}
    \includegraphics[width=\linewidth]{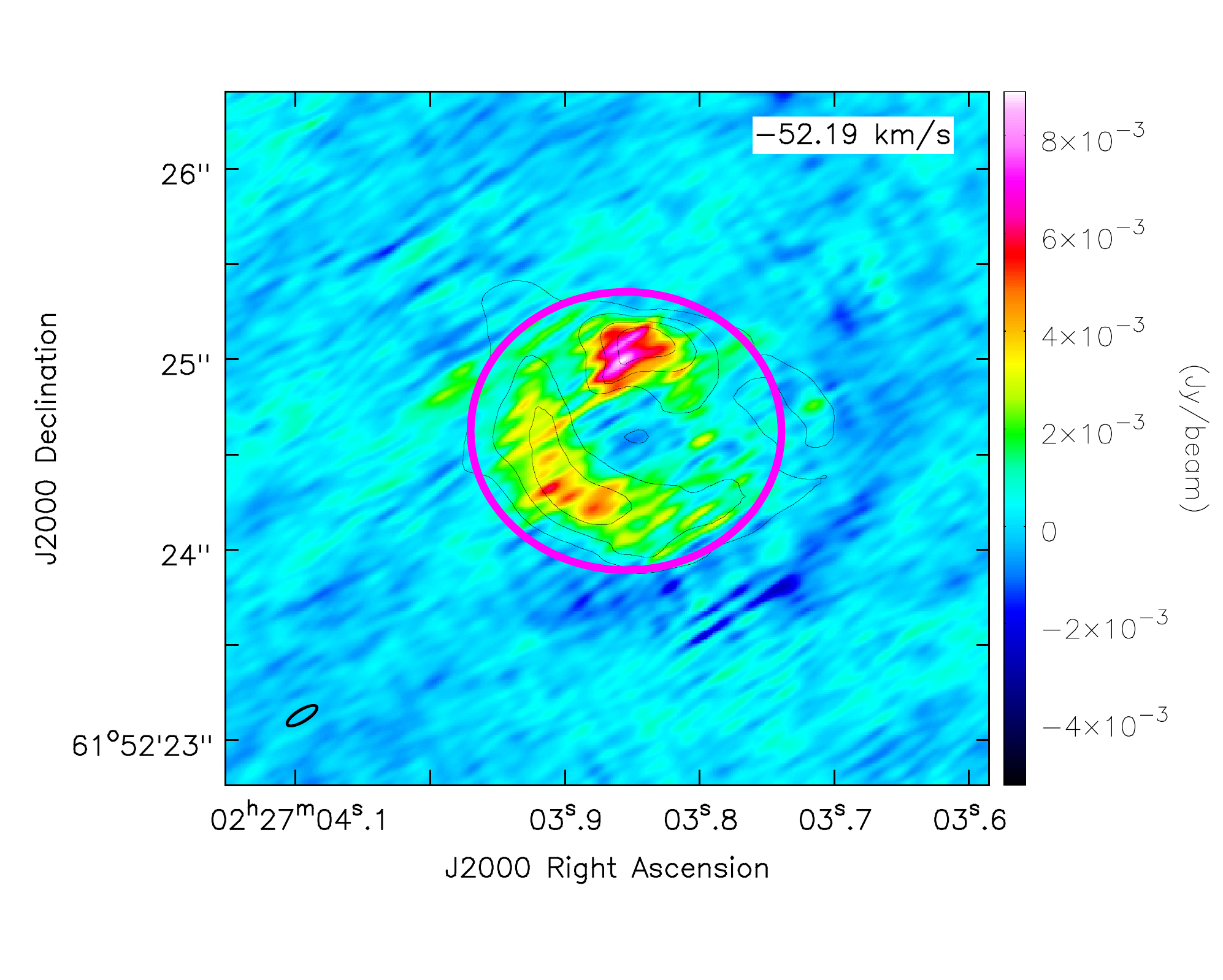}
  \end{minipage}
  \hfill
  \begin{minipage}[c]{0.48\linewidth}
    \vspace*{-1cm}
    \includegraphics[width=\linewidth]{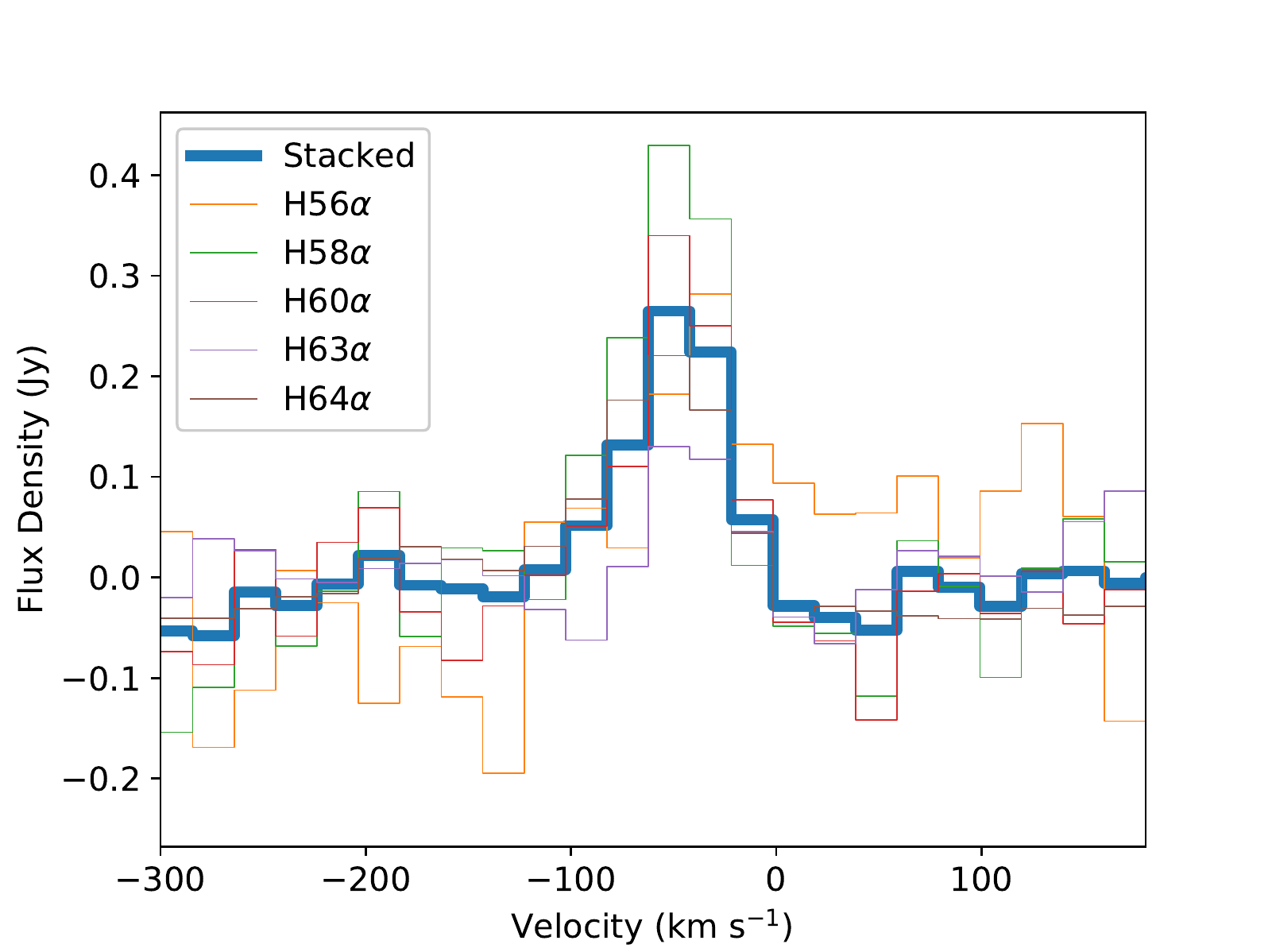}
  \end{minipage}
\caption{{\it Left Panel}: Contours show the radio continuum of W3(OH) at 27.5$\,$GHz (contours are 2, 4, 6, and 8 times the 3.7 mJy$\,$b$^{-1}$ RMS). The color image shows the peak channel of the stacked RRL cube (after continuum subtraction). The magenta ellipse shows the solid angle integrated to obtain the spectra shown in the right panel. {\it Right Panel}: Spectra of individual RRL transitions and the stacked spectrum (blue). }
\label{fig_W3OH_RRLs}
\end{figure}

In some cases, e.g., search for specific CH$_3$OH transitions, only Step 1 is needed, while in others, e.g., search for thermal lines from the same species, both steps are necessary. As a positive control test to check the procedure developed in this project (Appendix~\ref{sec:appendix}), we applied the script to continuum observations of the ultra-compact (UC) H{\small II} region W3(OH) (project number 14A-014; PI: L. F. Rodr\'{\i}guez). The observations were conducted in March 2014 with the VLA in A-configuration. The scheduling-block contained spectral windows between 18 and 37$\,$GHz; each spectral window had 64 channels, a bandwidth of 128$\,$MHz and a channel width of 2$\,$MHz. To make the test as automatic as possible, the data were only calibrated with the VLA pipeline without any additional flagging or self-calibration. The script selected spectral windows covering hydrogen RRL frequencies and created cubes using these spectral windows. Some spectral cubes had imaging artifacts and were subsequently excluded. The script then subtracted the radio-continuum in the image plane of the remaining cubes, re-sampled the spectra to a common channel width and synthesized beam, and stacked the cubes. Figure~\ref{fig_W3OH_RRLs} presents the result of the W3(OH) data; the left panel shows the radio continuum (in contours) and the peak channel of the stacked cube in colors. As expected, the peak emission of the stacked RRL data is coincident with the peak radio continuum of the UCH{\small II} region. The right-panel shows the spectra of all RRL transitions that were stacked to generate the blue spectrum. As is evident from the figure, the RRL signal is clearly visible in the stacked spectrum. Even though the W3(OH) observations significantly filtered out extended emission due to poorly sampled short spacing visibilities, particularly at the highest frequencies, and that the peak RRL flux densities are underestimated due to the large channel width of the continuum mode observations, Figure~\ref{fig_W3OH_RRLs} demonstrates that RRL lines can be detected from stacking of continuum-mode VLA observations. The potential of stacking spectral lines has been demonstrated in many previous studies (e.g., \citealt{Beuther_2016A&A...595A..32B}, \citealt{Jolly_2020MNRAS.499.3992J}, and references therein). In contrast, the novel aspect presented in this article is stacking of spectra from broadband data, thus, opening a new discovery space of low-velocity resolution spectral-line scans as by-product of VLA continuum observations.

The stacking approach used in this paper is similar to that of {\sc LINESTACKER} \citep{Jolly_2020MNRAS.499.3992J}, i.e., image-plane stacking based on {\sc CASA} tasks. We note that stacking in the $(u,v)$-plane is likely to provide better results than stacking in the image-plane, however, developing of such strategy in our code is beyond the objective of this work (see \citealt{Jolly_2020MNRAS.499.3992J} and references therein for a discussion of stacking in the $(u,v)$-plane). A practical difference between the tool developed in this project and {\sc LINESTACKER}  is that the script provided here requires calibrated MS files as inputs (not cubes, which are generated automatically by the script), while {\sc LINESTACKER} requires a list of pre-existing data cubes to be stacked. The strategy of using the calibrated MS file instead of pre-existing cubes facilitates the discovery process, as {\it a priori} knowledge of whether the rest frequency of a spectral line is in one of the SPWs of the MS file is not needed (the script searches for SPWs where spectral lines may be located). Moreover, creating spectral cubes of `continuum-mode' broadband SPWs is usually not done when creating continuum images, and therefore, cubes are not readily available to use  {\sc LINESTACKER} in typical reduction of continuum datasets. However, a caveat in the use of the script provided in this work is that the VLA calibration pipeline can flag (mask) very bright {\it bona-fide} spectral lines as radio frequency interference (RFI). Nevertheless, the automatic flagging is unlikely to remove most lines from continuum spectral windows given frequency dilution of narrow bright lines into broad channels. In case of doubt, users can remove the flags applied to the science sources of a calibrated dataset before running the script developed in this project.  

The script was applied to the 58 young high-mass star forming regions of the high-sensitivity VLA radio continuum survey carried out by \cite{Rosero_2016ApJS..227...25R}. We used the script to search for H$\alpha$RRLs, as well as all OH, CH$_3$OH and NH$_3$ transitions listed in Splatalogue\footnote{Spectroscopy information of all transitions in this article is from Splatalogue (\url{https://splatalogue.online/}).} up to energy levels between 500$\,$K and 10,000$\,$K depending on the molecular species and source (see below).

\section{Results and Discussion}\label{sec:results}

We report upper limits of excited OH transitions, detections of NH$_3$ and CH$_3$OH, and upper limits of RRLs. We cannot rule out that lines reported in this work are caused by transitions from other species (coincidental detections), however, given that the spectral density of bright molecular lines is low at cm wavelengths with respect to mm and sub-mm bands (e.g., ALMA) and the previous detections of these transitions toward other sources as reported in the literature (see discussion below), we consider that the possibility of misidentification of spectral lines is low. Moreover, several lines from similar excitation energies are detected for CH$_3$OH and NH$_3$ species, which suggests a correct identification of the lines. Nevertheless, high-spectral resolution observations are required to confirm the detections reported in this work and rule out misidentifications. Table~\ref{table_OH_limits} lists the OH transitions in the broadband spectral windows of the \cite{Rosero_2016ApJS..227...25R} sample, channel width and typical RMS. In the following sections, we discuss the NH$_3$, CH$_3$OH, and RRL results in more detail.

\newpage

\subsection{NH$_3$ Lines}\label{ssec:results_NH3}
\begin{figure}[ht]
\hspace*{0cm}
\vspace*{0cm}
\includegraphics[scale=0.37]{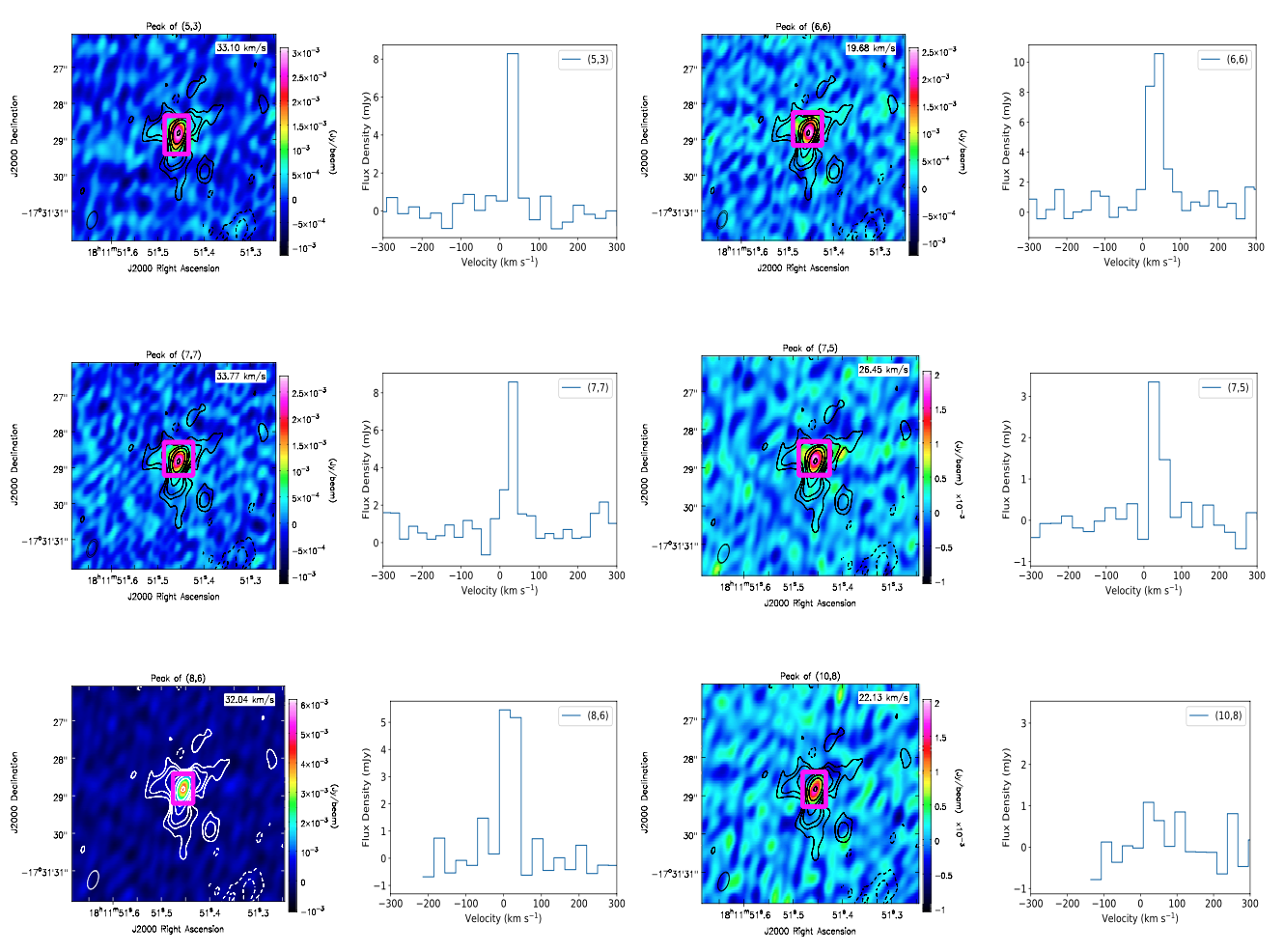}
\begin{minipage}[c]{1.0\linewidth}
\caption{NH$_3$ transitions included in the broadband continuum observations of IRAS 18089$-$1732A. Images of NH$_3$ peak intensity channels (colors) and 1.3$\,$cm radio continuum (contour levels are [$-$1.5, 3, 5, 7.5, 15, 25, 45, 95] times the 10 $\mu$Jy$\,$b$^{-1}$ RMS noise) are shown in left panels. Two very similar synthesized beams are shown superimposed on the lower left corner of the images, which represent the beam sizes of the continuum and NH$_3$ images. Corresponding spectra from the regions highlighted with magenta rectangles are shown in the right panels. Each region was first made to cover the 3$\sigma$ continuum source to search for detections, then its size was modified manually to match the emission region while minimizing contamination by noise or negative `sidelobes' near the detections. Note detections in all transitions with the exception of the (10,8) line, in which the signal is less than $\sim 3\sigma$. }
\end{minipage}
\label{fig_18089_NH3}
\end{figure}

\begin{figure}[h]
\hspace*{-1cm}
\vspace*{-2cm}
\includegraphics[scale=0.37]{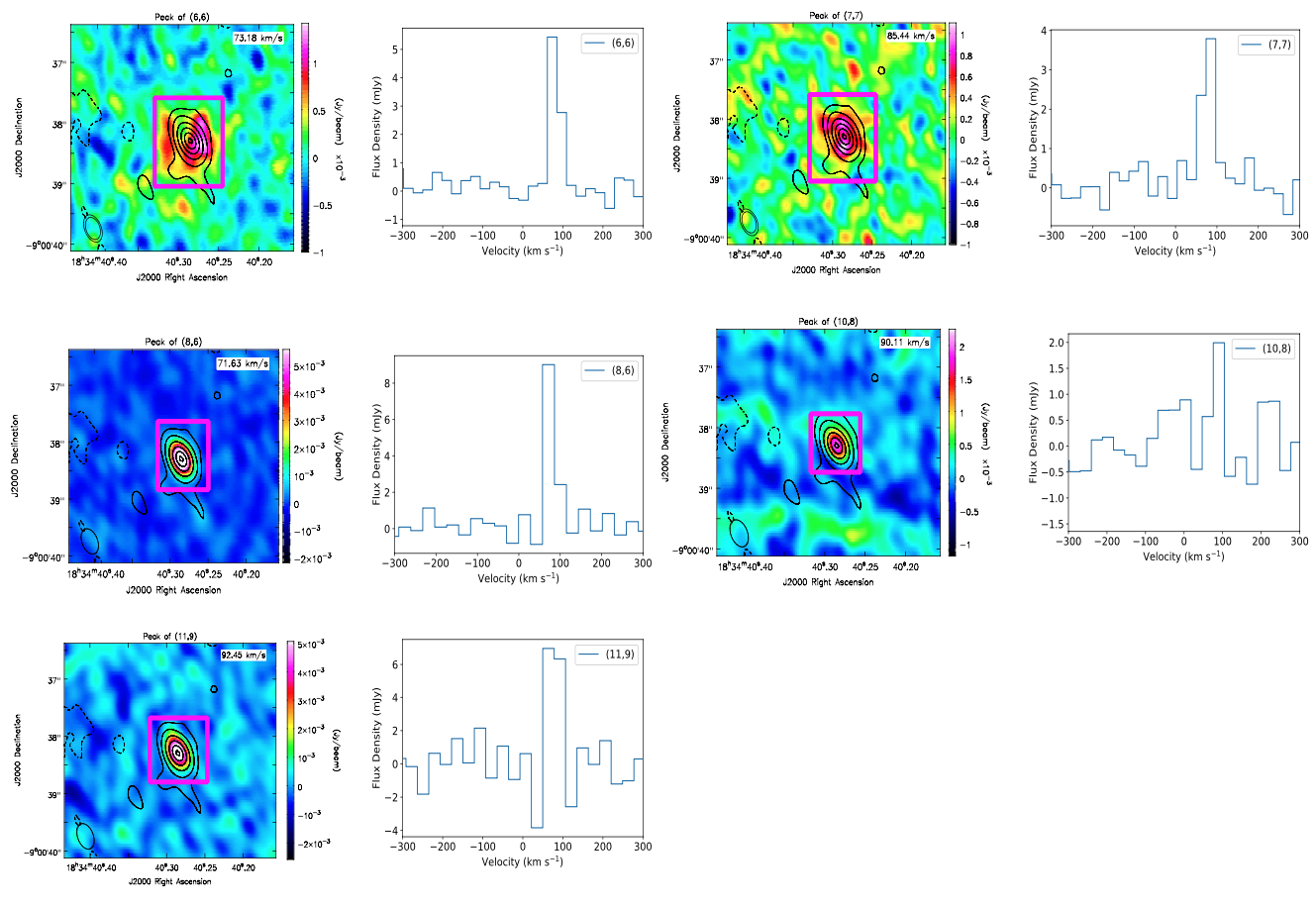}
\begin{minipage}[c]{1.0\linewidth}
\vspace*{3cm}
\caption{As Figure~\ref{fig_18089_NH3} but for G23.01$-$0.41A. The 1.3$\,$cm radio continuum contour levels are [$-$1.5, 3, 10, 25, 45, 65, 80] times the $9\mu$Jy$\,$b$^{-1}$ RMS noise. }
\end{minipage}
\label{fig_g2301_NH3}
\end{figure}

\begin{figure}[h]
\hspace*{-1cm}
\vspace*{-5cm}
\includegraphics[scale=0.40]{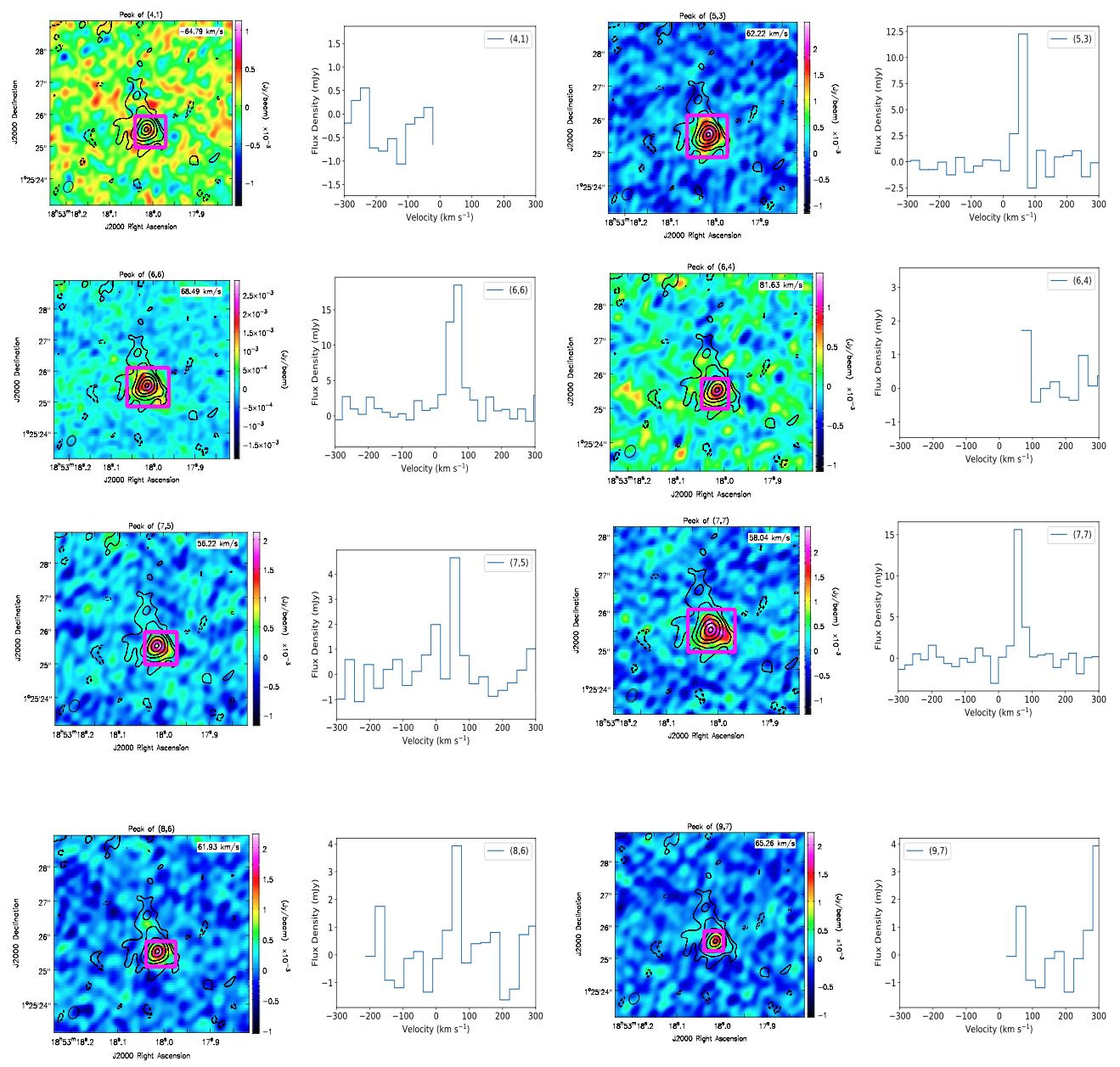}
\begin{minipage}[c]{1.0\linewidth}
\vspace*{6cm}
\caption{As Figure~\ref{fig_18089_NH3} but for G34.43+00.24mm1A. The 1.3$\,$cm radio continuum contour levels are [$-$2, 3, 6, 10, 20, 40, 60] times the $10\mu$Jy$\,$b$^{-1}$ RMS noise. The SPWs of the (4,1) and (6,4) transitions did not cover the systemic velocity of the region.}
\end{minipage}
\label{fig_g3443_NH3}
\end{figure}

\begin{figure}[h]
\includegraphics[width=1.0\textwidth]{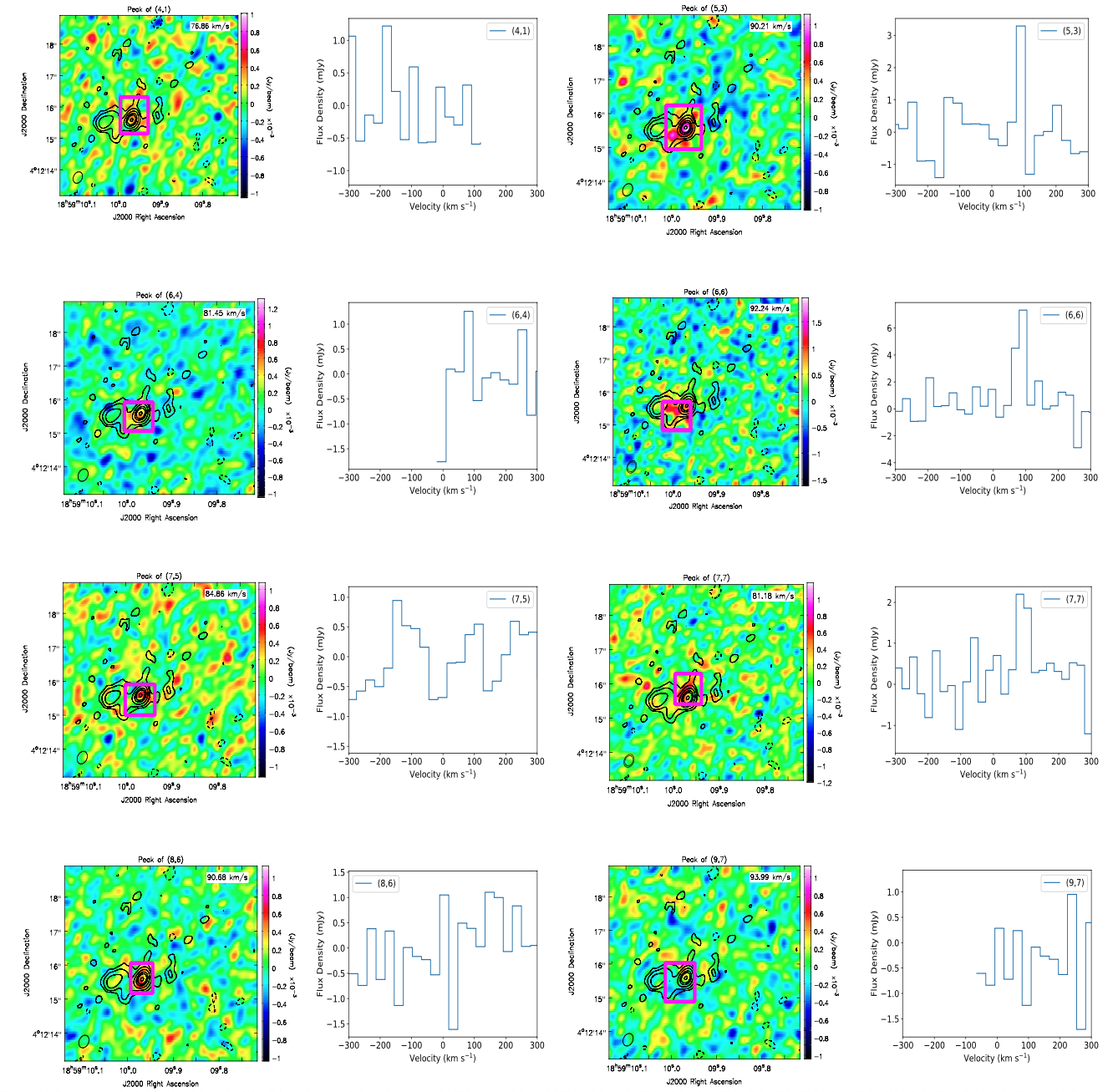}
\begin{minipage}[c]{1.0\linewidth}
\vspace*{1cm}
\caption{As Figure~\ref{fig_18089_NH3} but for IRAS$\,$18566+0408. The 1.3$\,$cm radio continuum contour levels are [$-$2, 3, 5, 9, 15, 20, 29] times the $7\mu$Jy$\,$b$^{-1}$ RMS noise. We report no detection of the (4,1), (6,4), (7,5), (8,6), and (9,7) transitions.}
\end{minipage}
\label{fig_18566_NH3}
\end{figure}

\begin{figure}[h]
\hspace*{-0.5cm}
\vspace*{-6cm}
\includegraphics[scale=0.37]{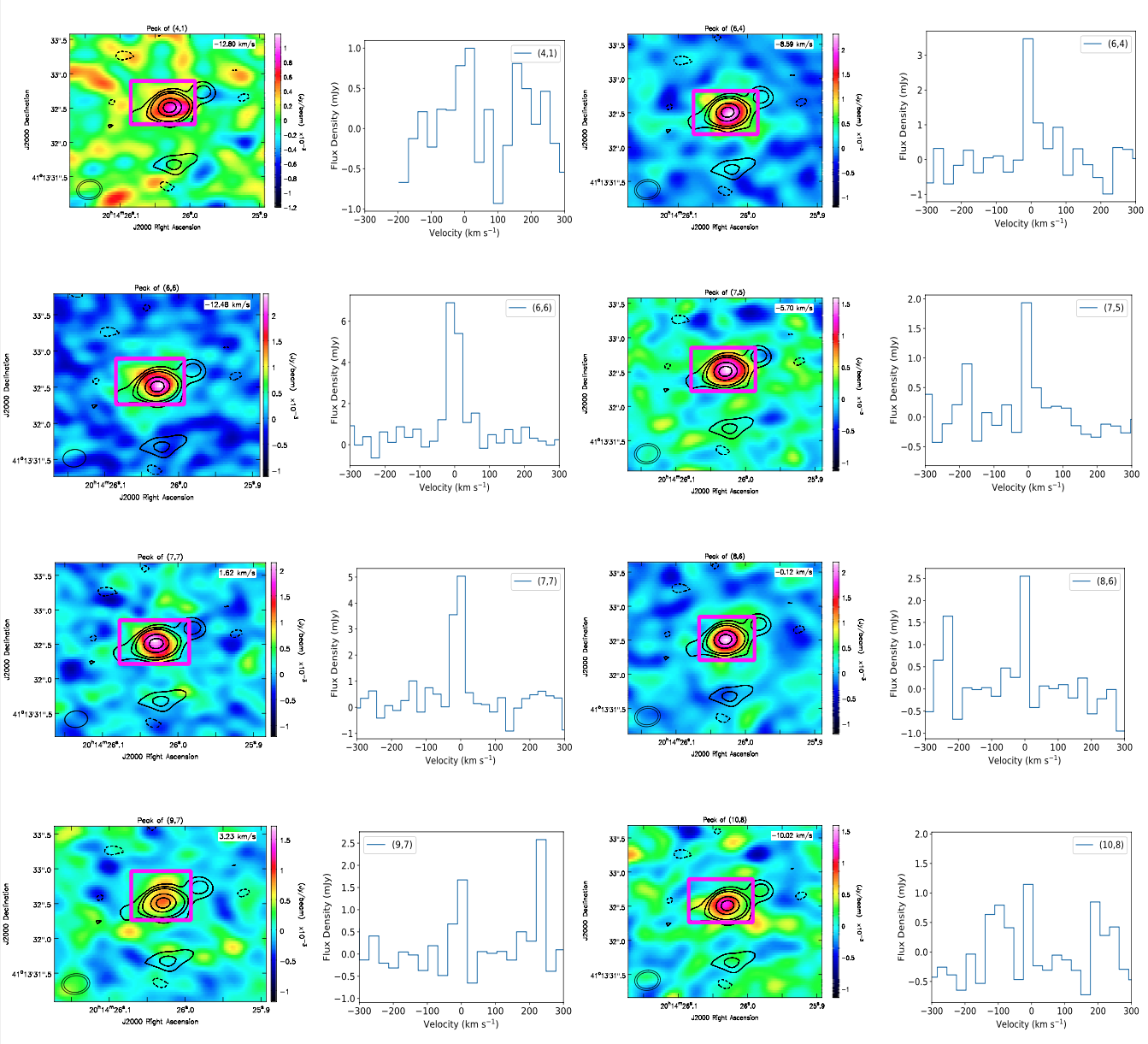}
\begin{minipage}[c]{1.0\linewidth}
\vspace*{7cm}
\caption{As Figure~\ref{fig_18089_NH3} but for IRAS$\,$20126+4104. The 1.3$\,$cm radio continuum contour levels are [$-$2, 3, 6, 10, 30, 50] times the $10\mu$Jy$\,$b$^{-1}$ RMS noise. We report no clear detection of the (4,1) and (10,8) transitions.}
\end{minipage}
\label{fig_IRAS2012_NH3}
\end{figure}

The script was used to find the NH$_3$ frequencies listed in Splatalogue that were serendipitously included (whether detected or not) in the SPWs of the \cite{Rosero_2016ApJS..227...25R} observations. We found that between 11 to 16 transitions were included in the SPWs when we searched for lines up to $E_l/k_B  = 10,000\,$K (minimum energy level corresponded to the transition (4,1) at $E_l/k_B  = 279\,$K and the maximum energy level corresponded to the transition (28,24) at  $E_l/k_B  =8359\,$K). Different sources have different number of transitions due to different tuning of the SPWs. We detected NH$_3$ emission lines toward five sources (Figures~\ref{fig_18089_NH3} to \ref{fig_IRAS2012_NH3}). Of them, NH$_3$ masers have been reported only toward IRAS$\,$20126+4104 but at other transitions [(3,3), (4,4); \citealt{Zhang_1999ApJ...527L.117Z}]. As shown in Table~\ref{table_NH3_parameters}, the detections include para ($K \ne 3n$) and ortho ($K = 3n$) states, from both metastable [(6,6), (7,7)] and non-metastable [(4,1), (5,3), (6,4), (8,6), (7,5), (9,7), (10,8)] transitions. For the five sources with NH$_3$ detections, 3 to 6 different NH$_3$ transitions were detected per source; in all cases the emission mostly coincides with the peak continuum in a single locus. The RMS of the spectra were typically smaller than 1$\,$mJy; the maximum peak flux density of 17.5$\,$mJy was found towards G34.43+00.24mm1A at 68\kms~(in a 24\kms~channel width).

\begin{figure}[h]
\includegraphics[scale=0.7,angle=-90]{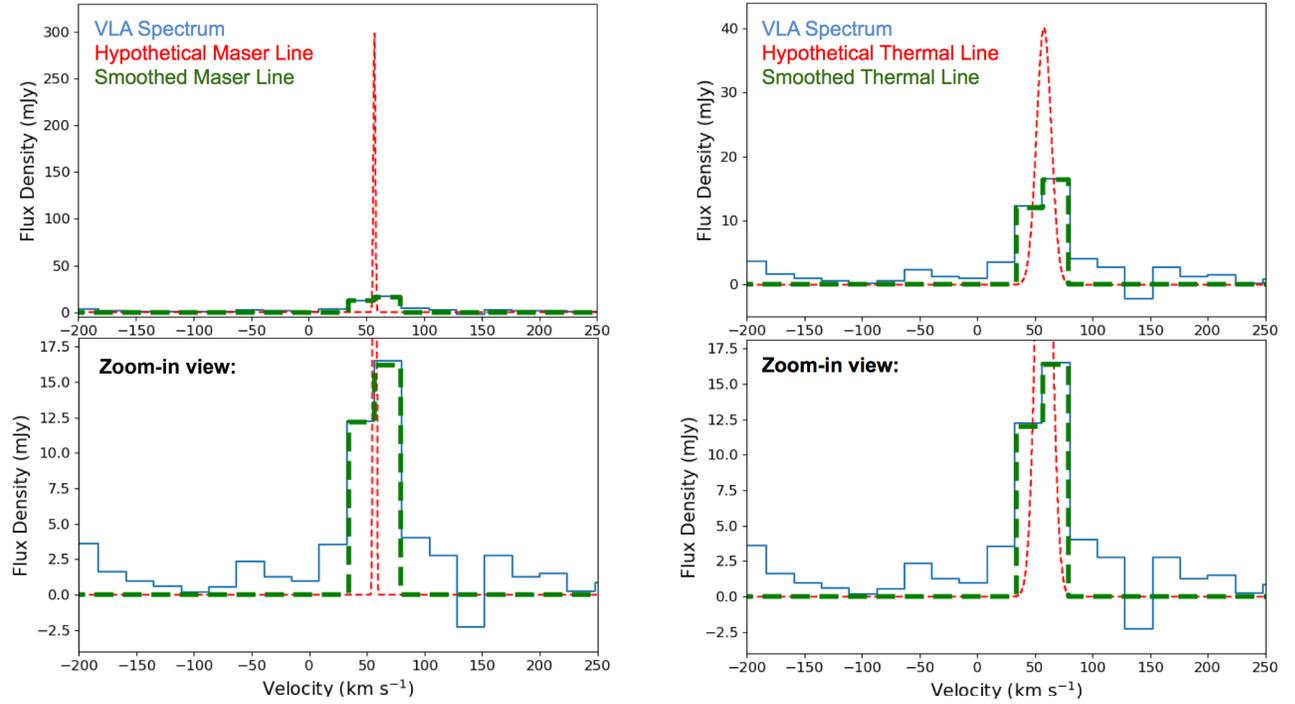}
\vspace*{-2cm}
\caption{Example of (6,6) detection toward G34.43+00.24mm1A (blue spectrum, all panels). The data can be fit equally well by a hypothetical strong and narrow maser line (red, left panels) or a hypothetical weaker and broader thermal line (red, right panels). In both cases, when the hypothetical lines are smoothed to the channel width of the data (green), the hypothetical lines match the data (zoom-in view, lower panels).}
\label{fig_NH3_fits}
\end{figure}

Given the large channel width of our data (23 to 29\kms~depending on the transition), we cannot use linewidths as proxy to determine whether the lines are due to a maser or thermal processes. For example, Figure~\ref{fig_NH3_fits} shows the (6,6) NH$_3$ spectrum toward G34.43+00.24mm1A (blue lines in all panels). The left panels show that a bright (300$\,$mJy) and narrow (2\kms~FWHM; red-dashed line) hypothetical maser can reproduce the observed spectrum when smoothed to our large channel width (green-dashed line). However, as shown in the right panels of Figure~\ref{fig_NH3_fits}, a hypothetical thermal line (40$\,$mJy peak and 15\kms~FHWM; red-dashed line) can also reproduce the observed data when smoothed to the channel width of the spectrum (green-dashed line). We note that as observed in the W51 region, thermal and maser lines of NH$_3$ can have values of peak flux density and FWHM similar to those used in Figure~\ref{fig_NH3_fits} (e.g., \citealt{Goddi_2015A&A...573A.109G}, \citealt{Henkel_2013A&A...549A..90H}).

We note that because of low spectral resolution, our brightness temperature lower limits ($T_b > 10\,$K) are too low to disentangle the maser or thermal nature of the lines, i.e., all NH$_3$ transitions reported in this work could be thermal. Nevertheless, a maser interpretation could be supported based on the angular size of the emission. For example, as shown in Figures~\ref{fig_18089_NH3} to \ref{fig_IRAS2012_NH3}, the NH$_3$ emission is mostly compact ($< 1$\arcsec) with emission cores smaller than 0.5\arcsec~in at least four of the sources. This is in contrast to more extended NH$_3$ thermal emission detected toward other sources, e.g., $\sim 4$\arcsec~angular size of (6,6) NH$_3$ emission in W51 IRS2 (\citealt{Goddi_2015A&A...573A.109G}; see also \citealt{Kraemer_1995ApJ...439L...9K}).

Based on the spatial distribution of our different detections (Figures~\ref{fig_18089_NH3} to \ref{fig_IRAS2012_NH3}), we find that the peaks of the NH$_3$ emission lines of metastable and non-metastable transitions are coincident with the peak radio continuum in most sources, although some positional offsets are observed, i.e., several transitions toward IRAS$\,$18566+0408 (Figure~\ref{fig_18566_NH3}). This behavior is similar to the (9,6) NH$_3$ masers toward Cepheus A and G34.26+0.15 \citep{Yan_2022A&A...659A...5Y}. Further high-spectral and angular resolution observations are needed to investigate the position and velocity offsets between different transitions, as, for example, \cite{Goddi_2015A&A...573A.109G} found that ortho [(6,6), (9,9)] and para [(7,7)] NH$_3$ masers in W51 show different spatial and velocity distributions.

High-excitation NH$_3$ masers have been reported toward tens of high-mass star forming regions (see \citealt{Wilson_1982A&A...110L..20W}; \citealt{Hofner_1994ApJ...429L..85H}; \citealt{Zhang_1999ApJ...527L.117Z}; \citealt{Walsh_2011MNRAS.416.1764W}; \citealt{Hoffman_2012ApJ...759...76H}; \citealt{Henkel_2013A&A...549A..90H}; \citealt{Mills_2018ApJ...869L..14M}; \citealt{Mei_2020ApJ...898..157M}; \citealt{Yan_2022A&A...659A...5Y}, and references therein), including non-metastable masers toward W51, W49, DR21, NGC$\,$7538, Cepheus A, G34.26+0.15, Sgr B2(N) and NGC$\,$6334 (with possible detection of the (11,9) and (8,6) transitions toward G19.61$-$0.23, \citealt{Walsh_2011MNRAS.416.1764W}) and metastable masers toward IRAS$\,$20126+4104, W33, Sgr$\,$B2 Main, the DR21 region (DR21(OH), DR21 H{\small II}), W51, G5.89$-$0.39, G9.62+0.19, NGC$\,$6334, in addition to possible detections toward Sgr$\,$B2 and G23.33$-$0.30, and a few active/starburst galaxies (IC$\,$342, \citealt{Gorski_2018ApJ...856..134G}, \citealt{Lebron_2011A&A...534A..56L}; NGC$\,$253, \citealt{Gorski_2017ApJ...842..124G}; NGC$\,$3079, \citealt{Miyamoto_2015PASJ...67....5M}). If high spectral resolution observations toward our detections confirm the maser interpretation, we will have added three new sources to the sample of high-mass star-forming regions with known NH$_3$ masers. Also, the (4,1) line would be the first detection of this maser transition in the ISM.

NH$_3$ masers are weaker with respect to other well known transitions of H$_2$O or CH$_3$OH. For example, the low-sensitivity ($\sim 2\,$Jy detection limit) HOPS survey resulted in only two new possible NH$_3$ masers out of a sample of hundreds of sources with H$_2$O masers \citep{Walsh_2011MNRAS.416.1764W}. As pointed out in the literature (e.g., \citealt{Goddi_2015A&A...573A.109G}; \citealt{Hoffman_2014ApJ...782...83H}) many questions on the nature of these masers remain unanswered, including whether the masers are tracers of outflows (\citealt{Zhang_1999ApJ...527L.117Z}, \citealt{Yan_2022A&A...659A...5Y}), disks/tori or quiescent material (e.g., see \citealt{Hoffman_2014ApJ...782...83H} for modeling of a (9,3) maser velocity gradient in terms of a Keplerian disk or a rotating torus).

\citealt{Hunter_2008ApJ...680.1271H}; \citealt{Beuther_2007A&A...466..989B}; \citealt{Hofner_1994ApJ...429L..85H}, \citealt{Zhang_Ho_1995ApJ...450L..63Z}; non-metastable transitions: \citealt{Yan_2022A&A...659A...5Y}; \citealt{Hoffman_2014ApJ...782...83H}; \citealt{Hoffman_2012ApJ...759...76H}; \citealt{Hoffman_2011ApJ...739L..15H}; \citealt{Walsh_2007MNRAS.382L..35W}; \citealt{Gaume_1993ApJ...417..645G}; \citealt{Pratap_1991ApJ...373L..13P}; \citealt{Wilson_1990A&A...229L...1W}). The importance of interferometric observations of these maser transitions is exemplified by W51-IRS2, where \cite{Goddi_2015A&A...573A.109G} found that the NH$_3$ masers do not originate from the prominent molecular outflow traced by SiO and H$_2$O masers but could be related to the outflow from a binary companion (at least the (7,7) transition; see also a similar interpretation of the (6,6) NH$_3$ maser in NGC$\,$6334I, \citealt{Beuther_2007A&A...466..989B}). In contrast, the NH$_3$ masers in NGC7538$\,$IRS1 could be associated with a rotating torus, where variability of the (3,3) $^{15}$NH$_3$ maser may be due to entrainment in the interface between the torus and the outflow (e.g., \citealt{Hoffman_2014ApJ...782...83H}, and references therein; see also interferometric observations of NH$_3$ (9,6) masers in Cep A and G34.26+0.25 reported by \citealt{Yan_2022A&A...659A...5Y}).

W51-IRS2 is one of the richest sources of NH$_3$ masers known (19 maser transitions have been reported; \citealt{Henkel_2013A&A...549A..90H}). The high detection rate of emission lines toward the five sources reported in this work (despite our low spectral resolution) indicates that these sources may also be rich in NH$_3$ masers. \cite{Goddi_2015A&A...573A.109G} reported the first imaging study of high-excitation NH$_3$ masers up to 850$\,$K in a high-mass star-forming region and \cite{Yan_2022A&A...659A...5Y} recently reported interferometric observations of the (9,6) line (1090$\,$K); if confirmed, we will have extended the imaging study of high-excitation NH$_3$ masers up to energies of 1220$\,$K above ground.

A common characteristic between the possible masers reported here and the ensemble of maser transitions in W51-IRS2 \citep{Henkel_2013A&A...549A..90H} is that the strongest line detected in each source corresponds to the only ortho-NH$_3$ metastable transition observed (Table~\ref{table_NH3_parameters}). As proposed by \cite{Henkel_2013A&A...549A..90H}, such strong deviation from LTE could be due to the greater statistical weights of ortho-states (resulting in greater column densities) and/or deviations caused by connection between the high excitation ortho-states to the $K=0$ level. Hence, future high angular and spectral resolution observations of the (3,3) NH$_3$ transition toward our sample could result in the detection of masers (e.g., \citealt{Zhang_Ho_1995ApJ...450L..63Z}), as is the case with IRAS$\,$20126+4104 \citep{Zhang_1999ApJ...527L.117Z}.

A particularly interesting aspect of the detections presented in this work is the lack of strong (greater than a few mJy at cm wavelengths, \citealt{Rosero_2016ApJS..227...25R}) radio continuum (in contrast to regions such as W51, Sgr$\,$B2 and W49). If confirmed, these NH$_3$ masers would be tracing very young phases of high-mass star formation before the development of ultracompact H{\small II} regions. \cite{Gorski_2017ApJ...842..124G} mention that NH$_3$ masers in starburst/active galaxies may originate from the interface of ionized outflows with surrounding molecular gas. Arguments supporting the association between NH$_3$ masers with the interface between ionized and molecular gas have been presented in the case of high-mass star-forming regions in the Galaxy (e.g., \citealt{Walsh_2011MNRAS.416.1764W}, \citealt{Hunter_2008ApJ...680.1271H}, \citealt{Beuther_2007A&A...466..989B}, \citealt{Kraemer_1995ApJ...439L...9K}), including the surfaces of hot expanding molecular shells \citep{Martin-Pintado_1999ApJ...519..667M}. If such an environment is also responsible for the putative NH$_3$ masers reported in this work, these masers may offer a window into the interaction between the expanding ionized jets and molecular envelopes, leading to momentum transfer into molecular outflows.

The pumping of NH$_3$ masers is still unclear as multiple mechanisms have been suggested depending on whether the masers are from metastable or non-metastable states (e.g., see \citealt{Mills_2018ApJ...869L..14M}). Proposed mechanisms include radiative excitation from a chance line overlap, collisional excitation of the ortho-states (e.g., \citealt{Yan_2022A&A...659A...5Y};  \citealt{Goddi_2015A&A...573A.109G}; \citealt{Zhang_1999ApJ...527L.117Z}; \citealt{Mangum_1994ApJ...428L..33M}; \citealt{Flower_1990MNRAS.244P...4F}; see also laboratory masers created via collisional excitation reported by \citealt{Willey_1995PhRvL..74.5216W}), and infrared pumping \citep{Madden_1986ApJ...300L..79M}. For example, \cite{Goddi_2015A&A...573A.109G} mention that, although collisional excitation is a viable mechanism for metastable ortho-NH$_3$ transitions, very high densities would be needed for the mechanism to operate for non-metastable states (see also \citealt{Henkel_2013A&A...549A..90H}). We note that our observations resulted in the detection of transitions with the same $K$ value [(6,6) and (8,6); (7,7) and (9,7)], which suggests that multiple transitions sharing the same $K$ state may be inverted; this should help constrain the physical conditions leading to the pumping (see \citealt{Brown_1991ApJ...378..445B}).

\subsection{CH$_3$OH Lines}\label{ssec:results_CH3OH}
\begin{figure}
\hspace*{1cm}
\includegraphics[scale=0.45]{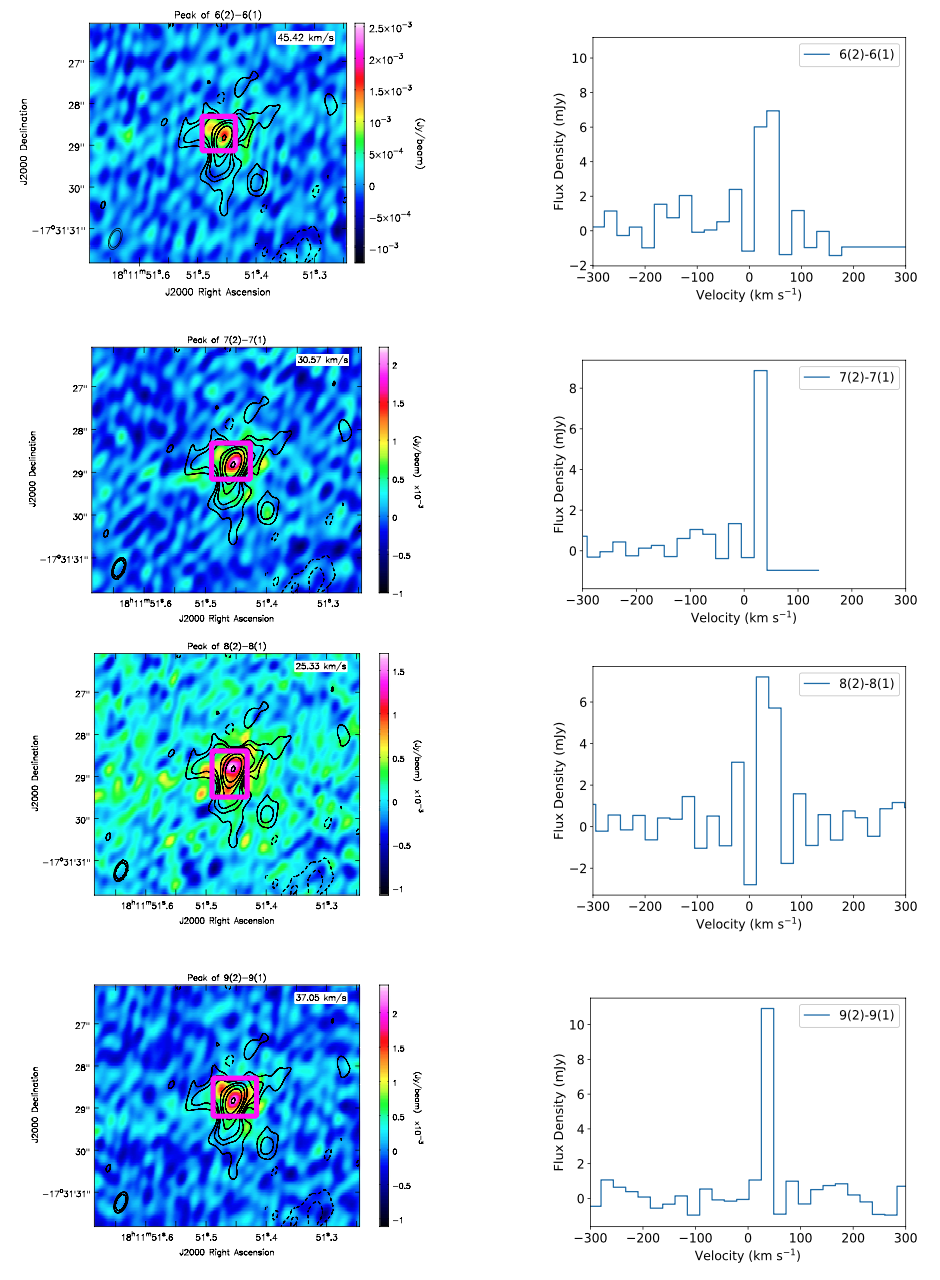}
\vspace*{0cm}
\caption{CH$_3$OH transitions detected in the broadband continuum observations of IRAS 18089$-$1732A. Images of CH$_3$OH peak intensity channels (colors) and 1.3$\,$cm radio continuum (contour levels are [$-$1.5, 3, 5, 7.5, 15, 25, 45, 95] times the 10$\,\mu$Jy$\,$b$^{-1}$ RMS noise) are shown in the left panels. The right panels show the corresponding spectra from the regions highlighted with magenta rectangles. As in the case of the NH$_3$ data, each region was first made to cover the 3$\sigma$ continuum source to search for detections, then its size was modified manually to match the emission region while minimizing contamination by noise or negative `sidelobes' near the detections. Class I 44$\,$GHz CH$_3$OH masers have been found in this region, but offset by $\sim 5\arcsec$~north of the continuum source \citep{Rodriguez-Garza_2017ApJS..233....4R}. }
\label{fig_peak_CH3OH_1}
\end{figure}

\begin{figure}[h]
\hspace*{1cm}
\includegraphics[scale=0.45]{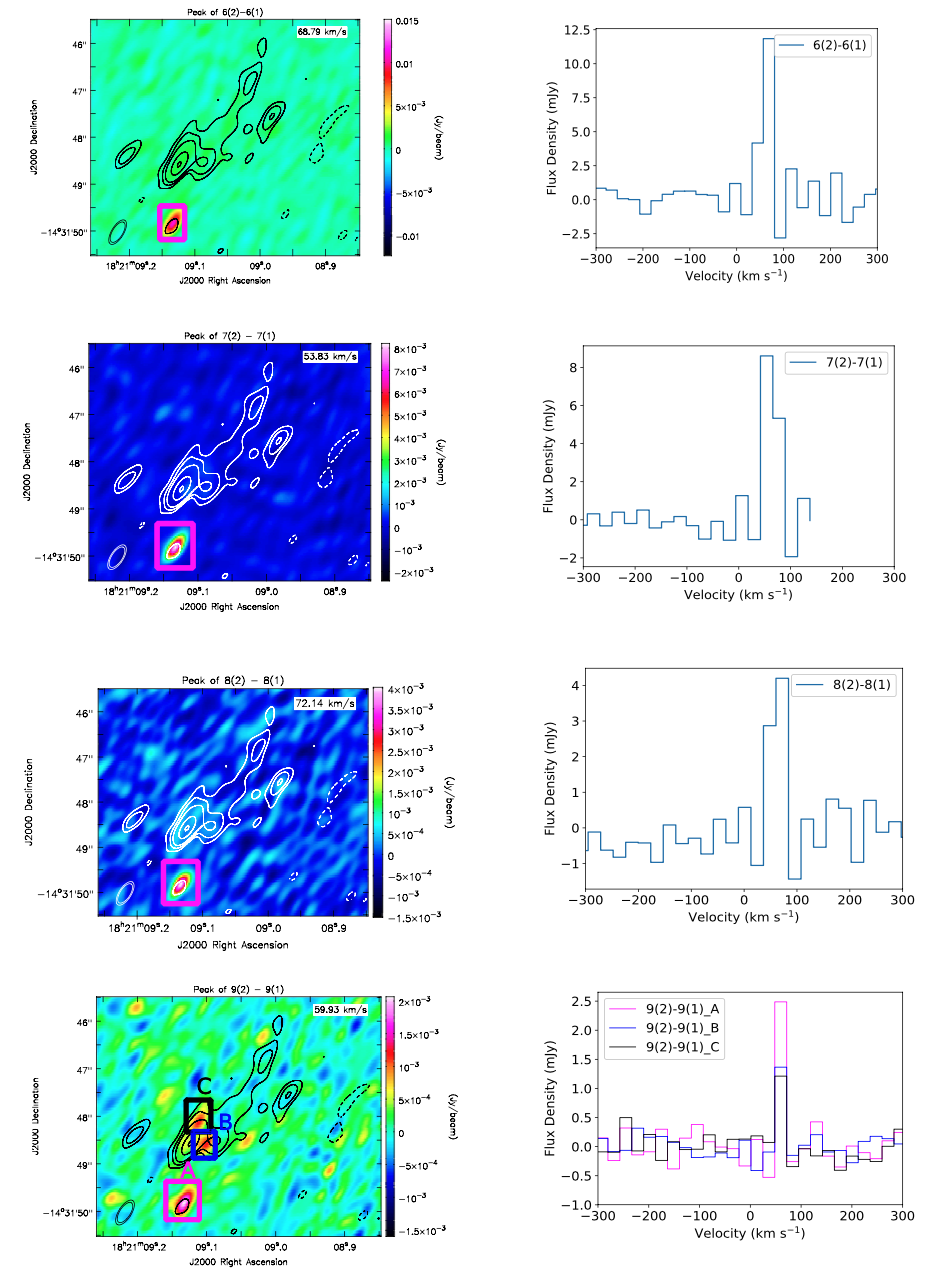}
\caption{As Figure \ref{fig_peak_CH3OH_1} but for IRAS 18182$-$1433. The 1.3$\,$cm radio continuum contour levels are [$-$2, 3, 5, 8, 20, 33] times the $9.5\mu$Jy$\,$b$^{-1}$ RMS noise. \cite{Rodriguez-Garza_2017ApJS..233....4R} reported the detection of a 44$\,$GHz CH$_3$OH maser toward source `A', which suggests that all detections toward this position are also Class I masers. Our detection of the 8(2)-8(1) line is consistent with the higher spectral resolution (albeit lower angular resolution) detection by \cite{Towner_2017ApJS..230...22T}. }
\label{fig_peak_CH3OH_2}
\end{figure}

As done with the NH$_3$ transitions, the script was used to find the CH$_3$OH frequencies listed in Splatalogue that were serendipitously included in the SPWs of the \cite{Rosero_2016ApJS..227...25R} observations (whether detected or not). We found that between 8 to 10 transitions were included in the SPWs up to energy levels of 500$\,$K (the transition with the minimum lower energy level was the 6(2)-6(1) E1 vt=0 line at  $E_l/k_B  =70\,$K, and the maximum lower energy level of a transition was the 8(2)-7(3) E1 vt=1 line at  $E_l/k_B  = 482\,$K). Different sources have different number of transitions due to different tuning of the SPWs. For sources with detections (see below), we explored whether other detections occurred at higher energy levels (we checked up to 5,000$\,$K) but none were found\footnote{Given the richness of CH$_3$OH spectra listed in Splatalogue, we found it impractical to search for spectral lines above 500$\,$K toward the complete sample. Cubes are generated for every transition, which requires significant computing time and auxiliary storage that could not be justified for sources with no detection of low energy transitions. This was particularly evident when sources with detection of low energy transitions resulted in non-detections above 500$\,$K.}. Out of the CH$_3$OH transitions serendipitously included in the SPWs, we report detection of CH$_3$OH emission at 25$\,$GHz toward ten sources. The detections correspond to the E-type $\Delta J = 0$, $\Delta K = 2 - 1$ series for $J$ values between 6 and 10 (see energy level diagram in \citealt{Leurini_2016A&A...592A..31L}); these transtions have been detected before in the interstellar medium (\citealt{Ladeyschiko_2019AJ....158..233L}, \citealt{Voronkov_2006MNRAS.373..411V}, \citealt{Towner_2017ApJS..230...22T}). Table~\ref{table_CH3OH} presents the line parameters for sources with detection in at least one transition. The RMS of the spectra were smaller than $\sim$2$\,$mJy; the maximum peak flux density of 19$\,$mJy was found towards G34.43+00.24mm1A at 61\kms. The mean RMS of the sources with detection in at least one transition is 0.54$\,$mJy, the mean flux density is 5.6$\,$mJy (24\kms~channel-width). Figures~\ref{fig_peak_CH3OH_1} to \ref{fig_peak_CH3OH_9} show the peak channel image and corresponding spectra of all sources with detection of at least one transition. In total, we detected emission of five different CH$_3$OH transitions, e.g., see the detections toward IRAS 19012+0536A (Table~\ref{table_CH3OH}). For some sources, some transitions were not included in the frequency band coverage of the observations, therefore, they are not listed in Table~\ref{table_CH3OH}, e.g., the 10(2)-10(1) transition was not observed toward IRAS$\,$18089$-$1732 A. In cases where the five transitions were observed toward a source, but not all were detected, we include detection limits in the table (e.g., IRAS$\,$18553+0414 A).

The 25$\,$GHz CH$_3$OH transitions detected in this work correspond to Class I masers when inverted (e.g., \citealt{Towner_2017ApJS..230...22T}). However, as in the case of NH$_3$ lines (Section~\ref{ssec:results_NH3}), the standard methods to establish the nature of the emission (thermal vs maser) using brightness temperature, linewidth, line ratios or spatial extent (e.g, \citealt{Towner_2017ApJS..230...22T}) are not conclusive in our case because the peak flux density of the lines is underestimated due to low spectral resolution, particularly in the case of narrow lines (see Figure~\ref{fig_NH3_fits}). Despite our high angular resolution ($\sim$0.4\arcsec), the average brightness temperature of CH$_3$OH lines in our sample is $\sim 50\,$K, which is too low to distinguish between maser and thermal emission, our channel width is too broad to determine linewidths. Extended CH$_3$OH emission (e.g., see the detections toward IRAS$\,$20126+4104, Figure~\ref{fig_peak_CH3OH_9}) could be due to the superposition of different maser spots, and a line ratio analysis is premature due to the underestimated flux density values as a result of spectral dilution. Nevertheless, we find it likely that our CH$_3$OH detections encompass both thermal and maser lines because:

\noindent {\it - Thermal: } 1. In some sources the CH$_3$OH detections are seen in all observed transitions toward the peak continuum, which is consistent with thermal emission (the detected transitions have similar, i.e., within a factor of $\sim 2$, energy above ground). 2. The energy levels of the transitions reported here (Table~\ref{table_CH3OH}) are between 70$\,$K and 150$\,$K, which could easily be achieved in hot molecular core environments of high-mass star forming regions (e.g., \citealt{Araya_2005ApJS..157..279A} and references therein).

\noindent {\it - Maser: } 1. While the CH$_3$OH lines in some sources trace the peak radio continuum (e.g., Figure~\ref{fig_peak_CH3OH_1}), in at least four cases (Figures~\ref{fig_peak_CH3OH_2}, \ref{fig_peak_CH3OH_5}, \ref{fig_peak_CH3OH_7}, \ref{fig_peak_CH3OH_8}) the CH$_3$OH emission is offset from the peak radio continuum as expected from Class I masers (e.g., \citealt{Kurtz_2004ApJS..155..149K}, \citealt{Cyganowski_2009ApJ...702.1615C}, \citealt{Araya_2009ApJ...698.1321A}). 2. Some transitions are found toward the same lines-of-sight (presumably the same physical locations) while other transitions are not. For example, the 9(2)-9(1) regions B and C in IRAS$\,$18182$-$1433 are not detected in the lower J transitions (Figure~\ref{fig_peak_CH3OH_2}), and the distribution of CH$_3$OH emission IRAS$\,$20126+4104 differs between transitions (Figure~\ref{fig_peak_CH3OH_9}), which is also expected in the case of masers (e.g., \citealt{Towner_2017ApJS..230...22T}).

To our knowledge, we report first detections in nine of the ten sources. The exception is IRAS$\,$18182$-$1433 (G16.59$-$0.05), which was also observed by \cite{Towner_2017ApJS..230...22T} with the VLA. They conducted higher spectral resolution (0.4\kms~channel width) but lower angular resolution ($\theta_{syn} \sim 1$\arcsec) observations compared to our data. They reported two 25$\,$GHz CH$_3$OH emission regions detected in the $3_2$, $5_2$, $8_2$, and $10_2$ transitions (see their Figure 1)\footnote{\cite{Towner_2017ApJS..230...22T} use a different nomenclature for the quantum transitions than the one used in this paper, e.g., their $8_2$ line corresponds to the $8(2)$-$8(1)$ line in our notation.}. The southern source, which they named G16.59$-$0.05\_b is coincident with our source A (Figure~\ref{fig_peak_CH3OH_2}), and was reported to have a narrow linewidth (0.8\kms) and peak flux density of the $5_2$ line of 320$\,$mJy, which implied a brightness temperature of 7200$\,$K, and therefore, was classified as a maser (the $5_2$ transition was not observed in our work). We note that such linewidth and flux density are very similar to those we used in Figure~\ref{fig_NH3_fits} to demonstrate the dichotomy between thermal and maser lines in our data. The other CH$_3$OH emission region reported by \cite{Towner_2017ApJS..230...22T} (G16.59$-$0.05\_a, which is coincident with our source B in Figure~\ref{fig_peak_CH3OH_2}) was classified as thermal because of its broader linewidth (2.8\kms) and low brightness temperature of the $5_2$ line (44$\,$K for a 36$\,$mJy line; see their Table 5); such a weak, narrow line would have been undetectable in our data. Therefore, our detection toward the same position (B), and our additional detection toward position C in a different transition (9(2)-9(1); Figure~\ref{fig_peak_CH3OH_2}) imply that the CH$_3$OH lines reported in this work toward B and C are also masers. The only transition we have in common with \cite{Towner_2017ApJS..230...22T} toward IRAS 18182$-$1433 is the $8_2$ line, for which we detect emission toward the region A (Figure~\ref{fig_peak_CH3OH_2}; G16.59$-$0.05\_b in \citealt{Towner_2017ApJS..230...22T}). They reported a peak flux density of 205$\,$mJy for this line; smoothing their line as illustrated in Figure~\ref{fig_NH3_fits} to a 24\kms~channel width results in a line that is consistent with our measurement of 4.3$\,$mJy within the RMS of our data. We note that the LSR velocities reported by \cite{Towner_2017ApJS..230...22T} toward the two CH$_3$OH emission regions in 18182$-$1433 (i.e., 58.6 and 61.4\kms) correspond very well with the LSR velocities we list in Table~\ref{table_CH3OH} (i.e., $\sim 60$\kms), which exemplifies the reliability of our method to find spectral lines from broadband VLA continuum observations (note that the channel width of our observations is $24$\kms).

Of the ten sources with 25$\,$GHz CH$_3$OH detections, six were previously observed by our group in \cite{Gomez-Ruiz_2016ApJS..222...18G} and \cite{Rodriguez-Garza_2017ApJS..233....4R} with the VLA in the 44$\,$GHz CH$_3$OH Class I maser transition (G23.01$-$0.41A, G34.43+00.24mm1A, IRAS 19012+0536A, and G53.25+00.04mm4A were not observed). All six sources harbor 44$\,$GHz CH$_3$OH masers, but as commonly observed with CH$_3$OH Class I transitions, the masers were mostly offset with respect to the central radio continuum source, in some cases by as much as $\sim 15$\arcsec~along aligned structures suggestive of shocks in outflows. Out of this sample of six sources, we found 44$\,$GHz CH$_3$OH masers coincident (within less than 0.5\arcsec~angular offset) with 25$\,$GHz CH$_3$OH detections toward IRAS 18182$-$1433 (region A in Figure~\ref{fig_peak_CH3OH_2}) and IRAS 18553+0414 A (Figure~\ref{fig_peak_CH3OH_5}). The spatial coincidence between the 25$\,$GHz CH$_3$OH detections reported in this work and the known 44$\,$GHz CH$_3$OH masers suggests that these specific detections correspond to Class I masers. In the case of the 25$\,$GHz CH$_3$OH detections found toward the radio continuum sources, the lines may correspond to thermal emission (e.g., see section 4.3 of \citealt{Towner_2017ApJS..230...22T}). However, in the two sources where 44$\,$GHz and 25$\,$GHz CH$_3$OH components are coincident, the angular separation between the masers and the radio continuum is less than $\sim 1$\arcsec, i.e., these are Class I masers found very close to young massive stellar objects. It is therefore possible that some others of our 25$\,$GHz CH$_3$OH detections may also be masers, but that the excitation of the 25$\,$GHz CH$_3$OH lines requires conditions found nearer massive YSO ($\lesssim 0.1\,$pc), while 44$\,$GHz masers are often found farther away from the continuum source (e.g., \citealt{Kurtz_2004ApJS..155..149K}, \citealt{Voronkov_2014MNRAS.439.2584V}). If this were the case, 25$\,$GHz CH$_3$OH Class I masers could preferentially be tracing shocked gas closer to the massive YSOs, and therefore, could trace the expansion of ionized jets, similar to the H$_2$O masers detected in W75N(B)-VLA2 \citep{Carrasco-Gonzalez_2015Sci...348..114C}. As pointed out by the numerical models of \cite{Leurini_2016A&A...592A..31L}, the masers from the 25$\,$GHz ladder are inverted at higher densities than other Class I masers (10$^6$ vs 10$^4\,$cm$^{-3}$) and therefore, may trace denser gas in shocks near massive YSOs. Further high-spectral resolution observations are needed to explore this possibility and to contrast the spatial distribution of possible masers in our sample with respect to the spatial distribution of 25$\,$GHz CH$_3$OH masers in other regions as reported in the literature (e.g., \citealt{Voronkov_2006MNRAS.373..411V}, \citealt{Towner_2017ApJS..230...22T}).

\subsection{Association with Ionized Jets}

We found that both NH$_3$ and CH$_3$OH detections are located very near the continuum sources (coincident in most cases, or within $2$\arcsec, which corresponds to a physical separation between $\sim 0.1 $ to $0.01\,$pc given the distances as listed in \citealt{Rosero_2016ApJS..227...25R}). Out of the ten regions with CH$_3$OH detection, five have been classified as ionized jets and three as ionized jet candidates, i.e., eight regions belong to the group of 25 sources listed as ionized jets or candidates in Tables~2 and 3 of \cite{Rosero_2019ApJ...880...99R}. The two exceptions are G34.43+00.24mm1A and G53.25+00.04mm4A, however, association with an ionized jet is also possible in at least one of them. In the case of G34.43+00.24mm1A (Figures~\ref{fig_g3443_NH3} and \ref{fig_peak_CH3OH_3}), the radio continuum is elongated and it is characterized by a rising spectral index $\alpha = +0.7\pm0.1$, which is consistent with dense ionized gas in a jet (e.g., see \citealt{Reynolds_1986ApJ...304..713R}). Moreover, \cite{Isequilla_2021A&A...649A.139I} reported the presence of multiple molecular outflows in this source; the main outflow being in the NE-SW direction, which is approximately colinear with the radio continuum elongation of the source (Figure~\ref{fig_g3443_NH3} contours; \citealt{Rosero_2016ApJS..227...25R})\footnote{G34.43+00.24mm1A would have been classified as an ionized jet in \cite{Rosero_2019ApJ...880...99R} if the \cite{Isequilla_2021A&A...649A.139I} results would have been known.}. The other source not classified as a jet or jet candidate by \cite{Rosero_2019ApJ...880...99R} is G53.25+00.04mm4A, which is unresolved ($\theta_{syn} \sim$0.4\arcsec, K-band, \citealt{Rosero_2016ApJS..227...25R}) and has a flat spectral index $\alpha = 0.1\pm0.1$. We note that multiple cm-continuum sources with different spectral indices can coexist within small physical areas in star forming regions (e.g., \citealt{Towner_2021ApJ...923..263T}, \citealt{Sanna_2019A&A...623L...3S}), therefore, higher sensitivity and angular resolution observations should be conducted to further investigate the nature of G53.25+00.04mm4A. 
We highlight that all five sources with NH$_3$ detection (Section~\ref{ssec:results_NH3}) were also detected in CH$_3$OH; it is likely that other regions with CH$_3$OH detection may also have NH$_3$ emission, albeit below our sensitivity levels. Our data therefore suggest that high-excitation NH$_3$ and CH$_3$OH lines can trace very young high-mass star forming regions harboring ionized jets.

\subsection{Radio Recombination Lines and Stacking}\label{ssec:results_rrls}

VLA radio continuum datasets of regions characterized by thermal ionized gas are ideal to search for RRLs as a by-product of broadband observations. However, our target sample consists of very young high-mass star forming regions and ionized jet candidates prior to the development of bright H{\small II} regions, therefore, any RRL would be very weak, as expected from ionized jets \citep{Anglada_2018A&ARv..26....3A}. We searched for $\Delta n = 1$ ($\alpha$) hydrogen transitions in the observed continuum frequencies, as they are expected to be the brightest RRLs (e.g., \citealt{Liu_2013AJ....146...80L})\footnote{Some carbon RRLs could be brighter than hydrogen RRLs, e.g., see spectrum of Mol$\,$160 in \cite{Araya_2007ApJS..170..152A}, however, hydrogen transitions would be brighter when smoothed to the large channel width of our data.}. A total of 8 RRL frequencies are within the observed spectral windows; Table~\ref{table_freqs} lists the specific transitions, channel widths and range of RMS values from the spectra of all sources.

We found no clear RRL detection in the stacked spectra. Figure~\ref{fig_RRL} shows examples of RRL spectra toward four sources. The left column shows C-band continuum detections in contours \citep{Rosero_2016ApJS..227...25R} superimposed to a channel image from the stacked C-band RRL cubes. The center and right columns of Figure~\ref{fig_RRL} show the C-band and K-band RRL spectra, respectively. The thin color lines in each panel show individual RRL transitions as identified in the insets, while the thick blue line shows the stacked spectrum per band. The RMS values of the stacked RRLs at C-band and K-band are listed in Table~\ref{table_RRLs}. As expected, the RMS of the stacked spectra is smaller than the RMS values of individual transitions. For regions with multiple continuum sources in the field (e.g., IRAS$\,$18566+0408, Figure~\ref{fig_RRL}), we show spectra of only one continuum source (or the combined spectra from a subset of sources in a region), but we searched for RRLs toward all continuum sources (not all spectra are shown). As explained in Section~\ref{sec:obs}, the individual RRL spectra were regrided to the same channel width, i.e., 133\kms~at C-band and 38\kms~at K-band, before stacking. The RMS values per source reported here are similar to state-of-the-art RRL surveys, albeit with broader channel width due to the nature of our VLA continuum data set. For example, one of the most sensitive large-scale RRL surveys to date is that of \cite{Liu_2019ApJS..240...14L}, which reported RMS values of $\sim 0.65\,$mJy with a 5.1\kms~channel width based on stacking of twelve hydrogen RRLs transitions from H163$\alpha$ to H174$\alpha$ (see Tables~\ref{table_freqs} and \ref{table_RRLs} for comparison).

We can obtain a statistical upper limit of RRL detectability in our sample by combining (stacking) all RRL cubes of all sources per band. To accomplish this, each cube was smoothed to the same spatial and spectral resolution, then sub-regions were obtained centered at the location of each known continuum source in all fields. Finally, all sub-region cubes were stacked. Such a stacking method assumes that the RRLs are aligned in velocity, which is a reasonable assumption given the large channel width of our data, especially at C-band (see above). Using this method, we obtained RMS values of 7.0$\,\mu$\JyB~at C-band and $17\,\mu$\JyB~at K-band (same channel widths as above). As it is well known from the radiometer equation, stacking $N$ different lines would decrease the RMS by $N^{-1/2}$ assuming the same sensitivity in all input spectra (e.g., \citealt{Anglada_2018A&ARv..26....3A}, \citealt{Jolly_2020MNRAS.499.3992J}). Figure~\ref{fig_ResultRC} shows the results of this stacking exercise. The figure shows the RMS values as a function of the number of stacked RRLs for C-band and K-band data. At an abscissa value of $10^0$ (one RRL transition), Figure~\ref{fig_ResultRC} shows the dispersion of RMS values for all RRLs (sources and transitions) per band. The RMS of the sample decreased when all RRL spectra per source were stacked (2 to 5 RRLs were included in the broadband continuum observations per band per source). When all spectra per band were stacked (all transitions and sources per band) then we obtained the right-most data point in both panels of Figure~\ref{fig_ResultRC}. We conclude that the RMS decreases with the number of stacked RRLs as expected, i.e., $RMS\, \propto N^{-1/2}$ (blue line, Figure~\ref{fig_ResultRC}). Although this trend is not expected to continue indefinitely, our results show that for VLA imaging the relation is valid at least through moderate samples of $N \sim 100$.

Despite the very low RMS obtained using this stacking method, we are still above the expected minimum flux density value for the combined RRL signal in our sample assuming optically thin emission ($S_{\nu,min}$). We can estimate this limit as follows: the weakest continuum source in our sample at C-band is UYSO1 B that has a peak continuum intensity of 15$\,\mu$\JyB~at 7.4$\,$GHz. Assuming optically thin free-free continuum as above, the RRL from this source would be $\sim 0.4\,\mu$\JyB. We can take this value as $S_{\nu,min}$ because all other continuum sources in the sample are brighter than UYSO1 B, and thus, the $\sim 0.4\,\mu$\JyB~value would be a lower limit of the expected optically thin emission in the stacked sample. This value is below the 7.0$\,\mu$\JyB~RMS at C-band of our stacked data, and thus, we cannot rule out optically thin free-free RRLs from the overall sample. Moreover, the continuum emission is likely not optically thin at C-band (see SEDs in \citealt{Rosero_2016ApJS..227...25R}), and the lines may be narrower than our channel width, which would result in an even lower value of the stacked RRL signal. The same argument applies to the K-band data based on the weakest continuum source in our sample (IRAS$\,$19266+1745C, 17$\,\mu$\JyB~at 20.9$\,$GHz; assuming a line-to-continuum ratio of 32\%), which results in an expected peak flux density of $\sim5\,\mu$\JyB, which is smaller than the RMS of the stacked spectrum ($17\,\mu$\JyB). We conclude that the stacking procedure works as expected (RMS decreases as $\sim N^{-1/2}$), and that we can rule out strong maser-like emission of RRLs in our overall sample, however, as expected, thermal RRL emission is too weak in our sample for a statistical detection.

The stacking method used in this project (instructions for installation and use of the script developed here are given in the Appendix~\ref{sec:appendix}) can be used for statistical detections toward a sample of brighter radio-continuum sources than those discussed here, as well as toward samples of weak continuum sources with the next generation of high-sensitivity interferometers such as the ngVLA.

\section{SUMMARY}
\label{sec:summary}

The broadband and multi-channel continuum observing mode of the VLA results in datasets that are effectively spectral scans of weak+broad or bright+narrow lines. We developed a script to search for spectral lines in calibrated continuum visibility files, and applied it to our radio continuum survey toward 58 young high-mass star forming regions \citep{Rosero_2016ApJS..227...25R}, which consisted of observations at C and K bands. We focused on the search for excited OH, NH$_3$, CH$_3$OH, and RRLs. In the case of the RRLs, the script stacked the different transitions per source to improve sensitivity. Given the low radio continuum intensity in our sample, our search for RRLs was done as a proof of concept to investigate whether the RMS decreases as expected from the radiometer equation and to search for unusually strong lines that could be indicative of non-thermal emission. We found that the stacking method decreases the RMS noise as expected, and report no detection of RRLs (3$\sigma$ detection limit of $\sim$0.2$\,$mJy at C-band with a 133\kms~channel width, and $\sim$0.4$\,$mJy at K-band with a channel width of 29\kms), within an angular size smaller than $\sim$1\arcsec.

We detected no excited OH lines in the sample ($3\sigma$ upper limit of $\sim$0.6$\,$mJy within an angular size smaller than 0.8\arcsec~and a channel width between 70 and 130\kms), but found multiple CH$_3$OH and NH$_3$ lines. Specifically, we report the first detection of several 25$\,$GHz CH$_3$OH transitions toward nine of the 58 regions in our sample; five sources have NH$_3$ detections. Due to the large channel width of our data ($\sim 70$ to 130\kms~at C-band; $\sim 20$ to 30\kms~at K-band) and the low flux density of the detections ($< 19\,$mJy), it is unclear whether the lines are due to a maser or thermal mechanism. High spectral resolution follow-up observations are required to investigate the nature of these detections. We found that both NH$_3$ and CH$_3$OH are located very near ionized jets or jet candidates (coincident in most cases, or within $2$\arcsec, $\lesssim 0.1\,$pc). Therefore, these transitions could be useful tracers of young massive stellar objects during the development of ionized jets. Particularly interesting is the case of CH$_3$OH detections that, if due to maser in origin, would imply Class I transitions tracing material very near ionized jets, which could reveal shocks due to the interaction of the jets with the molecular medium near the YSOs from which the outflows originate.

The script provided in this article to search for spectral lines and stacking RRLs from continuum VLA datasets should be especially applicable to continuum surveys toward extragalactic objects, in which low velocity resolution spectra would be appropriate for detection of broad lines (e.g., \citealt{Araya_2004ApJS..154..541A}, \citealt{Eisner_2019ApJ...882...95E}). This is a new discovery tool to explore the rich VLA archive of continuum observations and complement high-$z$ stacking efforts (e.g., see \citealt{Jolly_2020MNRAS.499.3992J} and references therein).

\acknowledgments

We would like to thank an anonymous referee for a detailed review that helped us to significantly improve this manuscript.  E.D.A. acknowledges support of WIU Distinguished Alumnus Frank Rodeffer to the WIU Physics Department and the WIU Astrophysics Research Laboratory, in particular student scholarships and computational resources. The National Radio Astronomy Observatory is a facility of the National Science Foundation operated
under cooperative agreement by Associated Universities, Inc. E.D.A. acknowledges partial support from NSF grant AST-1814063. P.H. acknowledges partial support from NSF grant AST-1814011. E.D.A. would like to thank WIU student Mr. Gbenga Miracle Adebisi for participating in the test of the code developed in this project. E.S.T. would like to thank the AAS for a FAMOUS grant to support participation in an AAS meeting. This research has made use of NASA's Astrophysics Data System. 

\software{CASA 5.1, 5.6, 6.2}

\bibliography{Spectral_Survey_VLA_jets}{}

\bibstyle{aasjournal}

\clearpage

\appendix
\section{appendix section}
\label{sec:appendix}

\subsection{Installation and Usage}
\label{sec:appendix_installation_and_usage}

The software developed in this work to search for spectral lines in continuum-mode VLA observations is made available to the community. The instructions for installation and use are:\\

1. Download source files using the {\tt git} command:\\

{\tt git clone \url{https://github.com/AstroLab-WIU/Spectral_Line_Search_and_Stack.git}}\\

This will create the folder {\tt Spectral\_Line\_Search\_and\_Stack} containing the Python (.py) scripts and other necessary files to execute the routine.\\

2. Open and edit the file {\tt parameters.txt}. This file contains multiple parameters that CASA requries to execute each task. The variables are separated in the following sections:

\begin{itemize}
 \item {\tt \#\#Path\#\#}

  - {\tt path\_to\_MS} = Reference path of where the visibility ({\tt *.ms}) files are located.
 
 \item {\tt \#\#Visibilities\#\#}

  - {\tt vis} = list of calibrated visibilities: could be one or multiple visibility files, e.g., {\tt [vis1.ms,vis2.ms, ...]}. Note, no {\tt ''} needed to list the ms files.
 
  - {\tt field =  ALL} or index (ID number) or field/source name, e.g., 2, or {\tt sourceA}. List of sources can be obtained with the CASA task {\tt listobs}. Note, no {\tt ''} needed in the input parameters.

 \item {\tt \#\#Frequency$\_$File\#\#}

   - {\tt molecule} = The name of the file that contains the frequencies of interest, e.g., {\tt CH3OH.tsv}. Frequency files are in the {\tt ./Species/} folder.

   - {\tt upper$\_$energy} = Upper limit of the energy level measured in Kelvin (i.e., maximum $E_l/k_B$) to search for lines with energy levels below this limit.

  \item {\tt \#\#Control$\_$Parameters\#\#}

   - {\tt generate$\_$cubes} = {\tt True}, the script will find the SPWs that contain rest frequencies listed in the {\tt Frequency$\_$File}, and create cubes accordingly. 

   - {\tt stack$\_$cubes} = {\tt True}, the script will run the stacking algorithm. In this case, the cubes from different spectral lines will be stacked to increase sensitivity.

  - {\tt generate$\_$stats} = {\tt True}, the script will display rms values of the cubes that were generated.

 \item {\tt \#\#Cube$\_$Gen\#\#}

    - Parameters of {\tt tclean} for imaging.\footnote{See: \url{https://casa.nrao.edu/casadocs/casa-6.1.0/global-task-list/task_tclean/parameters} for details.}

\end{itemize}

3. Execute script in a IPython CASA (version 5.1 or higher) terminal: {\tt execfile('main$\_$script.py')}.\\

4. The script will run multiple routines, execution time will depend on size of the file and workstation capabilities.

\subsection{Further Details}
\label{sec:appendix_details}

In this section we include additional details of the scripts provided as part of this work to benefit users who may want to modify them for specific applications: 

{\it - Scope of the Script:} The script designed in this work checks whether specific spectral line frequencies (as provided by the user in the {\tt Species} sub-directory) are included in the SPWs of a VLA MS file, if so, the script creates cubes of the identified SPWs using the rest frequency of the transitions. The script does not check if lines are detected or not, nor it identifies other potential lines that could be present within the SPWs, i.e., the script does not provide a list of spectral lines within a frequency range that could be detected within the SPW's bandpass. After the script is run and cubes are generated, the user has to check for detections, and for possible misidentification of lines due to broad channel widths. 

{\it - Continuum Subtraction: } As mentioned in Step 2 of  Section~\ref{sec:obs}, continuum subtraction is done as part of the stacking procedure. The code provided as part of this paper is designed for cases in which very weak spectral lines may be present, first because of spectral dilution due to the broad channel widths, but most importantly, because if stacking is needed, then the lines would have been weak (or undetectable) in individual SPWs to begin with. Therefore, the function {\tt chans\_rm\_continuum} in {\tt stacking\_module.py} obtains the RMS of each channel in a cube using the CASA task {\tt imstat} within an arbitrary box region with pixel coordinates '50,50,300,300'. The mean and standard deviation of the array of RMS values for each channel of the cube are calculated; channels with RMS values within the mean RMS $\pm$ three standard deviations of the RMS values are used to determine the baseline to remove the continuum using the CASA task {\tt imcontsub()}. This procedure removes from the bandpass used for baseline subtraction any channel with high RMS relative to the other channels in the cube due, for instance, to a bright maser or RFI. We note that whether or not the continuum source is within the arbitrary box used to calculate the RMS is not very relevant, as the method is based on deviations of the RMS per channel with respect to the mean RMS of all channels in the cube. We highlight that a user could modify the code to fit specific needs, e.g., use a different box region to obtain statistics, set a fixed range of channels for continuum subtraction, use a larger order polynomial to fit the baseline, etc.

{\it - CASA Tasks Used:} Here we provide a list of the main CASA tasks/functions used in the sequence of Step 1 and Step 2 mentioned in Section~\ref{sec:obs}: {\tt listobs()} (used to extract information of the dataset, including SPW frequencies); {\tt tclean()} (used to generate the spectral cubes); {\tt imstat()} (used to obtain statistical information of the images/cubes); {\tt imcontsub()} (used to remove the continuum in the image plane); {\tt imregid()} (to regrid all images to consistent spatial and spectral shapes for averaging); {\tt immath()} (to stack the cubes).

\clearpage

\begin{deluxetable}{lcccc}[h]
\tablecaption{OH Non-Detections \label{table_OH_limits}}
\tablewidth{0pt}
\tablehead{\colhead{Transition} & \colhead{Rest Freq.} & \colhead{E$_l/k_B$}  & \colhead{Channel Width} & \colhead{Average RMS} \\
\colhead{} & \colhead{(GHz)} & \colhead{(K)}& \colhead{(\kms)} & \colhead{(mJy/beam)}}
\startdata
OH $^2\Pi_{1/2}$ J=3/2 F=2$-$1 & 7.831955 & 270 &  77 & 0.15\\
OH $^2\Pi_{1/2}$ J=3/2 F=1$-$1 & 7.761743 & 270 &  77 & 0.15\\
OH $^2\Pi_{1/2}$ J=1/2 F=0$-$1 & 4.660241 & 182 & 126 & 0.20\\
\enddata
\tablenotetext{~}{Of the 58 pointings in the \cite{Rosero_2016ApJS..227...25R} sample, only 31 contained spectral windows with the rest frequencies of these OH transitions. The synthesized beams of SPWs with OH frequencies ranged from 0.2\arcsec to 0.8\arcsec.  The average 3$\sigma$ detection limit of the OH transition in the sample is $\sim$0.6$\,$mJy within an angular size smaller than 0.8\arcsec.}
\end{deluxetable}

\begin{deluxetable}{lccc ccc ccc cc}[h]
\tabletypesize{\scriptsize}
\tablecaption{NH$_3$ Line Parameters  \label{table_NH3_parameters}}
\tablewidth{0pt}
\tablehead{
\colhead{Source}& \multicolumn{2}{c}{J2000 Coordinates}   & \colhead{Rest Freq.} & \colhead{E$_l$/$k_B$}  &\colhead{Line} &\colhead{RMS} & \colhead{S$_\nu$} & \colhead{V$_{LSR}$} & \colhead{Linewidth}  & \colhead{$\int $S$_\nu dv$} \\
\cline{2-3}
\colhead{}   & \multicolumn{1}{c}{R.A.} & \multicolumn{1}{c}{Decl.}       & \colhead{(GHz)} & \colhead{(K)}  &\colhead{(J,K)} &\colhead{(mJy)} & \colhead{(mJy)}  &  \colhead{(\kms)}   & \colhead{(\kms)}&  \colhead{(mJy~\kms)}}
\startdata
IRAS$\,$18089$-$1732A & 18:11:51.455 & $-$17:31:28.76  & 21.285275 & 379 &    (5,3) &	 0.50 &  8.29   & 33(28)   &     28 &    233(14) \\
                                     & 18:11:51.455 & $-$17:31:28.76 & 25.056025  & 407 &    (6,6) &  0.38 &  10.6   & 44(24)   &     48 &    453(13) \\
	 	 	             & 18:11:51.455 & $-$17:31:28.76 & 25.715182 &	537 &    (7,7) &	 0.83 &  8.58   & 34(23)   &     23 &    200(19) \\
	 	 	             & 18:11:51.455 & $-$17:31:28.76 & 20.804830 &	665 &    (7,5) &	 0.41 &  3.30   & 26(29)   &     58 &    136(17) \\
	 	 	             & 18:11:51.455 & $-$17:31:28.76 & 20.719221 &	834 &    (8,6) &	 0.70 &  5.45   &  3(29)   &     58 &    307(28) \\
	 	 	             &                      &                        & 20.852527 &  1226 & (10,8) & 0.54 & $<1.62$ & \nodata  & \nodata&	 \nodata \\
G23.01$-$0.41A             & 18:34:40.288 & $-$9:00:38.16  & 25.056025 &	407 &    (6,6) &	 0.28 &  5.36   &  73(24)  &     48 &    193(10) \\
	 	 	             & 18:34:40.288 & $-$9:00:38.16  & 25.715182 &	537 &    (7,7) &	 0.38 &  3.84   &  85(23)  &     47 &    145(11) \\
	 	 	             & 18:34:40.288 & $-$9:00:38.16  & 20.719221 &	834 &    (8,6) &	 0.48 &  8.95   &  72(29)  &     58 &    327(20) \\
	 	                     & 18:34:40.288 & $-$9:00:38.16  & 20.852527 &  1226 &  (10,8) &  0.60 &  2.11   &  90(29)  &     29 &     61(17) \\
	 	 	             & 18:34:40.288 & $-$9:00:38.16  & 21.070739 &  1449 &  (11,9) & 1.01 &  6.74   &  64(28)  &     57 &    366(40) \\ 
G34.43+00.24mm1A      &                     &                         & 21.134311 &   279 &     (4,1) & 0.76 & \nodata &  \nodata & \nodata&    \nodata \\
	 	 	             & 18:53:18.009 &  1:25:25.51      & 21.285275 &   379 &      (5,3) & 0.51 & 12.3    &  62(28)  &     56 &    421(20) \\
	 	 	             & 18:53:18.009 &  1:25:25.51      & 25.056025 &   407 &	(6,6) & 0.95 & 17.5    &  68(24)  &     72 &    786(39) \\
	 	 	             &                     &                        & 20.994617 &   513 &	(6,4) &  0.80 & \nodata & \nodata  & \nodata&	 \nodata\\
	 	                     & 18:53:18.009 &  1:25:25.51      & 25.715182 &   537 &	(7,7) &  1.10 &  15.6   &   58(23) &	 47 &    461(35) \\
	 	 	             & 18:53:18.009 &  1:25:25.51      & 20.804830 &   665 &	(7,5) &  0.41 &  4.65   &   56(29) &     29 &   134(12) \\
	 	 	             & 18:53:18.009 &  1:25:25.51      & 20.719221 &   834 &	(8,6) &  0.44 &  3.87   &   62(29) &     29 &    112(13) \\
	 	 	             & 18:53:18.009 &  1:25:25.51      & 20.735452 &  1021 &	(9,7) &  0.50 &  1.75   &   65(29) &     29 &    51(14) \\ 
IRAS$\,$18566+0408      &                     &                        & 21.134311 &   279 &	(4,1) &  0.56 & $<1.68$ & \nodata  & \nodata&	 \nodata \\
	 	 	             & 18:59:09.964 &  4:12:15.55      & 21.285275 &   379 &	(5,3) &  0.68 &  3.23   &   90(28) &     28 &    91(19) \\ 
                                     & 18:59:09.973 &  4:12:15.18      & 25.056025 &   406 &	(6,6) &  0.89 & 6.99    &   92(24) &     48 &    267(30) \\ 
	 	 	             &                      &                        & 20.994617 &   513 &	(6,4) &  0.78 &$<2.34$  &  \nodata & \nodata&	 \nodata \\
                                     & 18:59:09.980 &  4:12:15.76      & 25.715182 &   537 &        (7,7) & 0.59 & 2.14    &   81(23) &     47 &    92(20) \\ 
                                     &                     &                        & 20.804830 &   665 &	(7,5) &  0.51 & $<1.53$ &  \nodata & \nodata&	 \nodata \\
                                     &                     &                        & 20.719221 &   834 &	(8,6) &  0.78 &$<2.34$  & \nodata  & \nodata&	 \nodata \\
                                     &                     &                        & 20.735452 &  1021 &	(9,7) &  0.48 &$<1.44$  & \nodata  & \nodata&	 \nodata \\
IRAS$\,$20126+4104      &                     &                       & 21.134311 &   279 &	(4,1) &  0.50 &$<1.50$  &  \nodata & \nodata&	 \nodata \\
                                     & 20:14:26.017 &  41:13:32.49   & 25.056025 &   406 &	(6,6) &  0.37 & 6.90    & $-$12(24)&     48 &    294(13) \\
                                     & 20:14:26.017 &  41:13:32.49   & 20.994617 &   513 &	(6,4) &  0.46 & 3.53    & $-$9(29) &     29 &    101(13) \\
                                     & 20:14:26.017 &  41:13:32.49   & 25.715182 &   537 &	(7,7) &  0.52 & 5.04    &    2(23) &     47 &    200(17) \\
                                     & 20:14:26.017 &  41:13:32.49   & 20.804830 &   665 &	(7,5) &  0.48 & 1.91    & $-$6(29) &     29 &     55(14) \\
                                     & 20:14:26.017 &  41:13:32.49   & 20.719221 &   834 &	(8,6) &  0.34 & 2.48    & $-$0(29) &     29 &     72(10) \\ 
                                     & 20:14:26.017 &  41:13:32.49   & 20.735452 &  1021 &	(9,7) &  0.38 & 1.66    & 3(29)    &     29 &     48(11) \\
                                     &                     &                       & 20.852527 &  1226 &      (10,8) &  0.43 &$<1.29$  & \nodata  & \nodata& \nodata    \\ 
\enddata
\tablenotetext{~}{Line parameters of NH$_3$ data sorted by $E_l/k_B$. Transitions listed with no data (\nodata) in the $S_\nu$ column are those for which the SPW did not cover the systemic velocity of the region, and therefore, the non-detection is unreliable. For sources with non-detections that include the systemic velocity of the source, then we report 3$\times$RMS flux density upper limits.  Channel width is reported as uncertainty in the peak channel ($V_{LSR}$) column. The velocity range above 3$\times$RMS is listed as Linewidth. The linewidth of the detections is one or two channels (see $V_{LSR}$ uncertainty with respect to the linewidth value, therefore, the linewidths should be consider upper limits, e.g., see Figure~\ref{fig_NH3_fits}). The $\int $S$_\nu dv$ uncertainty is RMS $\times$ channel width $\times$ square root of the number of channels in the line.}

\end{deluxetable}

\newpage

\begin{deluxetable}{lcc ccc ccc cc}[h]
\tabletypesize{\scriptsize}
\tablecaption{CH$_3$OH Line Parameters \label{table_CH3OH}}
\tablewidth{0pt}
\tablehead{
\colhead{Source} &	 \multicolumn{2}{c}{J2000 Coordinates}    &	 \colhead{Rest Freq.} &	 \colhead{E$_l$/$k_B$}  &	\colhead{Line} &	\colhead{RMS} &	 \colhead{S$_\nu$} &	 \colhead{V$_{LSR}$} &	 \colhead{Linewidth}  &	 \colhead{$\int $S$_\nu dv$} \\
\cline{2-3}
\colhead{}   &	 \multicolumn{1}{c}{R.A.} &	 \multicolumn{1}{c}{Decl.}               &	 \colhead{(GHz)} &	 \colhead{(K)}  &	\colhead{(J,K)} &	\colhead{(mJy)} &	 \colhead{(mJy)}  &	  \colhead{(\kms)}   &	 \colhead{(\kms)}&	  \colhead{(mJy~\kms)} }
\startdata
IRAS$\,$18089$-$1732A & 18:11:51.455 & $-$17:31:28.76 & 25.018176 &    70 &  6(2)-6(1) &	1.07    & 6.58    & 45(24)   & 48    & 293(36)   \\
                     & 18:11:51.455 & $-$17:31:28.76 & 25.124932 &    86 &  7(2)-7(1) &	0.55    & 8.62    & 31(24)   & 24    & 206(13)   \\
                     & 18:11:51.455 & $-$17:31:28.76 & 25.294483 &   105 &  8(2)-8(1) &	0.57    & 7.09    & 25(24)   & 47    & 301(19)   \\
                     & 18:11:51.455 & $-$17:31:28.76 & 25.541467 &   126 &  9(2)-9(1) &	0.60    & 10.9    & 37(23)   & 23    & 257(14)   \\
IRAS$\,$18182$-$1433 & 18:21:09.134 & $-$14:31:49.83 & 25.018176 &    70 &  6(2)-6(1) &	0.62    & 11.7    & 69(24)   & 48    & 374(21)   \\
                     & 18:21:09.134 & $-$14:31:49.83 & 25.124932 &    86 &  7(2)-7(1) &	0.48    & 8.71    & 54(24)   & 48    & 337(16)   \\
                     & 18:21:09.134 & $-$14:31:49.83 & 25.294483 &   105 &  8(2)-8(1) &	0.42    & 4.32    & 72(24)   & 47    & 173(14)   \\
~~~A                 & 18:21:09.134 & $-$14:31:49.83 & 25.541467 &   126 &  9(2)-9(1) &	0.24    & 2.55    & 60(23)   & 23    & 60(6)     \\
~~~B                 & 18:21:09.100 & $-$14:31:48.57 & 25.541467 &   126 &  9(2)-9(1) &	0.16    & 1.35    & 60(23)   & 23    & 32(4)     \\
~~~C                 & 18:21:09.112 & $-$14:31:48.13 & 25.541467 &   126 &  9(2)-9(1) &	0.24    & 1.23    & 60(23)   & 23    & 29(6)     \\
G23.01$-$0.41A       & 18:34:40.288 & $-$9:00:38.16  & 25.018176 &    70 &  6(2)-6(1) &	0.45    & 4.41    & 75(24)   & 24    & 106(11)   \\
                     &                      &                        & 25.124932 &    86 &  7(2)-7(1) & 0.45 &\nodata  &  \nodata &\nodata& \nodata\\
                     & 18:34:40.288 & $-$9:00:38.16  & 25.294483 &   105 &  8(2)-8(1) &	0.76    & 5.21    & 78(24)   & 24    & 123(18)   \\
                     & 18:34:40.288 & $-$9:00:38.16  & 25.541467 &   126 &  9(2)-9(1) &	0.63    & 3.79    & 89(23)   & 47    & 164(21)  \\
G34.43+00.24mm1A     & 18:53:18.009 &    1:25:25.51  & 25.018176 &    70 &  6(2)-6(1) &	1.97    & 15.11   & 46(24)   & 48    & 720(67)  \\
                     & 18:53:18.009 &    1:25:25.51  & 25.124932 &    86 &  7(2)-7(1) &	0.83    & 16.05   & 55(24)   & 48    & 495(28)  \\
                     & 18:53:18.009 &    1:25:25.51  & 25.294483 &   105 &  8(2)-8(1) &	0.87    & 15.04   & 50(24)   & 47    & 538(29)  \\
                     & 18:53:18.009 &    1:25:25.51  & 25.541467 &   126 &  9(2)-9(1) &	0.85    & 19.11   & 61(23)   & 23    & 449(20)  \\
                     & 18:53:18.009 &    1:25:25.51  & 25.878337 &   149 & 10(2)-10(1)& 1.22    & 11.65   & 48(23)   & 46    & 486(40)  \\    
IRAS$\,$18517+0437A  & 18:54:14.244 &    4:41:40.87  & 25.018176 &    70 &  6(2)-6(1) & 1.30    &  5.57   & 46(24)   & 24    & 134(31) \\
                     & 18:54:14.244 &    4.41.40.87  & 25.124932 &    86 &  7(2)-7(1) & 0.67    & 2.82    & 55(24)   & 24    & 67(16)   \\
                     & 18:54:14.244 &    4.41.40.87  & 25.294483 &   105 &  8(2)-8(1) & 0.55    & 3.82    & 50(24)   & 24    & 91(13)   \\
                     & 18:54:14.244 &    4.41.40.87  & 25.541467 &   126 &  9(2)-9(1) & 0.68    & 2.70    & 38(23)   & 23    & 63(16)   \\
                     & 18:54:14.244 &    4.41.40.87  & 25.878337 &   149 & 10(2)-10(1)& 0.72    & 3.49    & 48(23)   & 23    & 81(17)   \\
IRAS$\,$18553+0414A  & 18:57:53.349 &    4:18:16.73  & 25.018176 &    70 &  6(2)-6(1) & 0.22    & 2.93    & 22(24)   & 48    & 106(7)   \\
                     & 18:57:53.349 &    4:18:16.73  & 25.124932 &    86 &  7(2)-7(1) & 0.12    & 1.93    & 7(24)    & 48    & 60(4)   \\
                     & 18:57:53.349 &    4:18:16.73  & 25.294483 &   105 &  8(2)-8(1) & 0.16    & 0.96    & 2(24)    & 47    & 42(5)   \\
                     &              &                & 25.541467 &   126 &  9(2)-9(1) & 0.11    & $<$0.33 & \nodata  &\nodata& \nodata \\
                     &              &                & 25.878337 &   149 & 10(2)-10(1)& 0.13    & $<$0.39 & \nodata  &\nodata& \nodata \\
IRAS$\,$18566+0408   & 18:59:09.995 &    4:12:15.29  & 25.018176 &    70 &  6(2)-6(1) &	0.60    & 3.55    & 94(24)   & 48    & 136(20)   \\
                     & 18:59:09.995 &    4:12:15.29  & 25.124932 &    86 &  7(2)-7(1) &	0.54    & 2.18    & 79(24)   & 24    &  52(13) \\
                     & 18:59:09.995 &    4:12:15.29  & 25.294483 &   105 &  8(2)-8(1) &	0.46    & 2.05    & 97(24)   & 24    & 49(11)   \\
                     & 18:59:09.995 &    4:12:15.29  & 25.541467 &   126 &  9(2)-9(1) &	0.46    & 2.69    & 85(23)   & 23    & 63(11) \\
                     & 18:59:09.995 &    4:12:15.29  & 25.878337 &   149 & 10(2)-10(1)& 0.43    & 2.59    & 94(23)   & 23    & 60(10)   \\
IRAS$\,$19012+0536A  & 19:03:45.374 &    5:40:41.33  & 25.018176 &    70 &  6(2)-6(1) & 0.33    & 4.96    & 70(24)   & 24    & 119(8)   \\
                     & 19:03:45.374 &    5:40:41.33  & 25.124932 &    86 &  7(2)-7(1) & 0.38    & 3.89    & 79(24)   & 48    & 156(13)   \\
                     & 19:03:45.374 &    5:40:41.33  & 25.294483 &   105 &  8(2)-8(1) & 0.22    & 4.52    & 73(24)   & 47    & 146(7)   \\
                     & 19:03:45.374 &    5:40:41.33  & 25.541467 &   126 &  9(2)-9(1) & 0.31    & 2.46    & 61(23)   & 47    & 80(10)   \\
                     & 19:03:45.374 &    5:40:41.33  & 25.878337 &   149 & 10(2)-10(1)& 0.31    & 1.78    & 71(23)   & 23    & 41(7)   \\
G53.25+00.04$\,$mm4A  &              &                & 25.018176 &    70 &  6(2)-6(1) & 0.18    & $<$0.54 &\nodata   &\nodata& \nodata \\
                     & 19:29:34.191 &   18:01:38.45  & 25.124932 &    86 &  7(2)-7(1) & 0.22    & 0.97    & 6(24)    & 24    & 23(5)   \\
IRAS$\,$20126+4104   & 20:14:26.040 &   41:13:32.54  & 25.018176 &    70 &  6(2)-6(1) & 0.63    & 3.51    & $-$11(24)& 24    & 84(15) \\
                     & 20:14:26.040 &   41:13:32.54  & 25.124932 &    86 &  7(2)-7(1) & 0.54    & 5.22    & $-$2(24) & 24    & 125(13)   \\
                     & 20:14:26.040 &   41:13:32.54  & 25.294483 &   105 &  8(2)-8(1) & 0.47    & 5.74    & $-$7(24) & 24    & 136(11)   \\
                     & 20:14:26.040 &   41:13:32.54  & 25.541467 &   126 &  9(2)-9(1) & 0.66    & 5.60    & 5(23)    & 47    & 197(22)   \\
\enddata
\tablenotetext{~}{Line parameters of CH$_3$OH data sorted by $E_l/k_B$. Transitions listed with no data (\nodata) in the $S_\nu$ column are those for which the SPW did not cover the systemic velocity of the region, and therefore, the non-detection is unreliable. For sources with non-detections that include the systemic velocity of the source, then we report 3$\times$RMS flux density upper limits. Channel width is reported as uncertainty in the peak channel ($V_{LSR}$) column. The velocity range above 3$\times$RMS is listed as Linewidth. The linewidth of the detections is one or two channels (see $V_{LSR}$ uncertainty with respect to the linewidth value, therefore, the linewidths should be consider upper limits, e.g., see Figure~\ref{fig_NH3_fits}). The $\int $S$_\nu dv$ uncertainty is RMS $\times$ channel width $\times$ square root of the number of channels in the line.}
\end{deluxetable}

\clearpage

\begin{deluxetable}{lccc}[ht]
\tablecaption{RRLs in Spectral Windows \label{table_freqs}}
\tablewidth{0pt}
\tablehead{
\colhead{RRLs } &	 \colhead{Rest Freq.} &	 \colhead{Channel width} &	 \colhead{RMS Range}  \\
\colhead{}      &	 \colhead{(GHz)}      &	 \colhead{(\kms)}       &	 \colhead{(mJy/beam)} }
\startdata
H63$\alpha$  & 25.686280 & 23  &  $ 0.13$ - $0.37$ \\
H67$\alpha$  & 21.384784 & 28  &  $ 0.10$ - $0.38$ \\
H68$\alpha$  & 20.461765 & 29  &  $ 0.09$ - $0.38$ \\
H96$\alpha$  & 7.31829     & 82  &  $ 0.08$ - $0.19$ \\
H97$\alpha$  & 7.09541     & 85  &  $ 0.08$ - $0.18$ \\
H110$\alpha$ & 4.87415   & 123 &  $ 0.09$ - $0.16$ \\
H111$\alpha$ & 4.74418   & 126 &  $ 0.09$ - $0.16$ \\
H113$\alpha$ & 4.49776   & 133 &  $ 0.10$ - $0.32$ \\
\enddata
\tablenotetext{~}{ List of hydrogen RRL $\alpha$-transitions in the SPWs of our data. The last column shows the range of RMS values across all sources for which the systemic velocity was covered in the SPWs. The RMS values were measured for the cube of each source and transition using the CASA task {\tt imstat()}.}
\end{deluxetable}

\startlongtable
\begin{deluxetable}{lcc}
\tabletypesize{\scriptsize}
\tablecaption{RRL Detections Limits\label{table_RRLs}}
\tablewidth{0pt}
\tablehead{\colhead{Source} &\colhead{RMS C-Band} &\colhead{RMS K-Band}\\
\colhead{}       &\colhead{(mJy/beam)} &\colhead{(mJy/beam)}}
\startdata
LDN1657A$-$3     & $0.04$ & $0.29$\\
UYSO1 A          & $0.04$ & $0.21$ \\ 
UYSO1 B          & $0.05$ & $0.21$ \\
G11.11$-$012P1   & $0.07$ & $0.14$\\
IRAS 18089$-$1732A   & $0.06$ & $0.18$\\
IRAS 18151$-$1208AB & $0.06$ & $0.10$\\
IRAS 18151$-$1208C   & $0.07$ & $0.10$\\
IRAS 18182$-$1433     & $0.05$ &  $0.12$\\
IRDC 18223$-$3        & $0.06$ & $0.15$\\
IRAS 18264$-$1152     & $0.06$ & $0.09$\\
G23.01$-$0.41A   & $0.08$ & $0.11$\\
IRAS 18337$-$0743A   & $0.06$ &$0.08$ \\
IRAS 18345$-$0641A   & $0.08$ & $0.18$\\
IRAS 18437$-$0216A   & $0.09$ & \nodata\\
IRAS 18440$-$0148A   & $0.05$ & $0.16$\\
IRAS 18470$-$0044A   & $0.05$ & $0.17$\\
IRAS 18470$-$0044     & $0.05$ & $0.17$\\
G34.43+00.24mm1A      & $0.06$ & $0.12$\\
G34.43+00.24mm2       & $0.05$ & $0.10$\\
IRAS 18517+0437A     & $0.06$ & $0.16$\\
IRAS 18517+0437B     & $0.05$ & $0.16$\\
IRAS 18521+0134A     & $0.05$ &  $0.20$\\
IRAS 18521+0134B     & $0.05$ & $0.19$\\
G35.39$-$00.33mm2A   & $0.04$ & $0.19$\\
IRAS 18553+0414A       & $0.06$ & $0.15$\\
IRAS 18566+0408       & $0.04$ & $0.17$\\
IRAS 19012+0536A     & $0.07$ & $0.16$\\
IRAS 19035+0641       & $0.04$ & $0.15$\\
IRAS 19266+1745AB       & $0.05$ & $0.05$\\
IRAS 19266+1745C      & $0.05$ & $0.06$\\
G53.11+00.05mm2A        & $0.08$ & $0.06$\\
G53.25+00.04 mm2A     & $0.05$ & $0.12$\\
G53.25+00.04 mm4A     & $0.07$ & $0.11$\\
IRAS 19282+1814A       & $0.06$ & \nodata\\
IRAS 19411+2306A       & $0.07$ & $0.10$\\
IRAS 19413+2332A     & $0.07$ & $0.09$\\
IRAS 20126+4104     & $0.07$ & $0.15$\\
IRAS 20126+4104C    & $0.08$ & $0.14$\\
IRAS 20216+4107A     & $0.08$ & $0.11$\\
IRAS 20293+3952     & $0.04$ & $0.07$\\
IRAS 20293+3952D   & $0.05$ & $0.07$\\
IRAS 20293+3952E   & $0.05$ & $0.08$\\
IRAS 20343+4129A   & $0.06$ & $0.11$\\
IRAS 20343+4129B   & $0.06$ & $0.10$\\
\enddata
\tablenotetext{}{RMS values measured from stacked spectral lines. The synthesized beams of SPWs with RRL frequencies ranged from 0.2\arcsec to 1.1\arcsec.  The average 3$\sigma$ detection limit of hydrogen RRL $\alpha$-transitions in the sample is 0.2$\,$mJy at C-band, and 0.4$\,$mJy at K-band, within an angular size smaller than $\sim 1$\arcsec~synthesized beam.}
\end{deluxetable}

\clearpage

\clearpage

\clearpage

\begin{figure}[h]
\hspace*{1cm}
\includegraphics[scale=0.45]{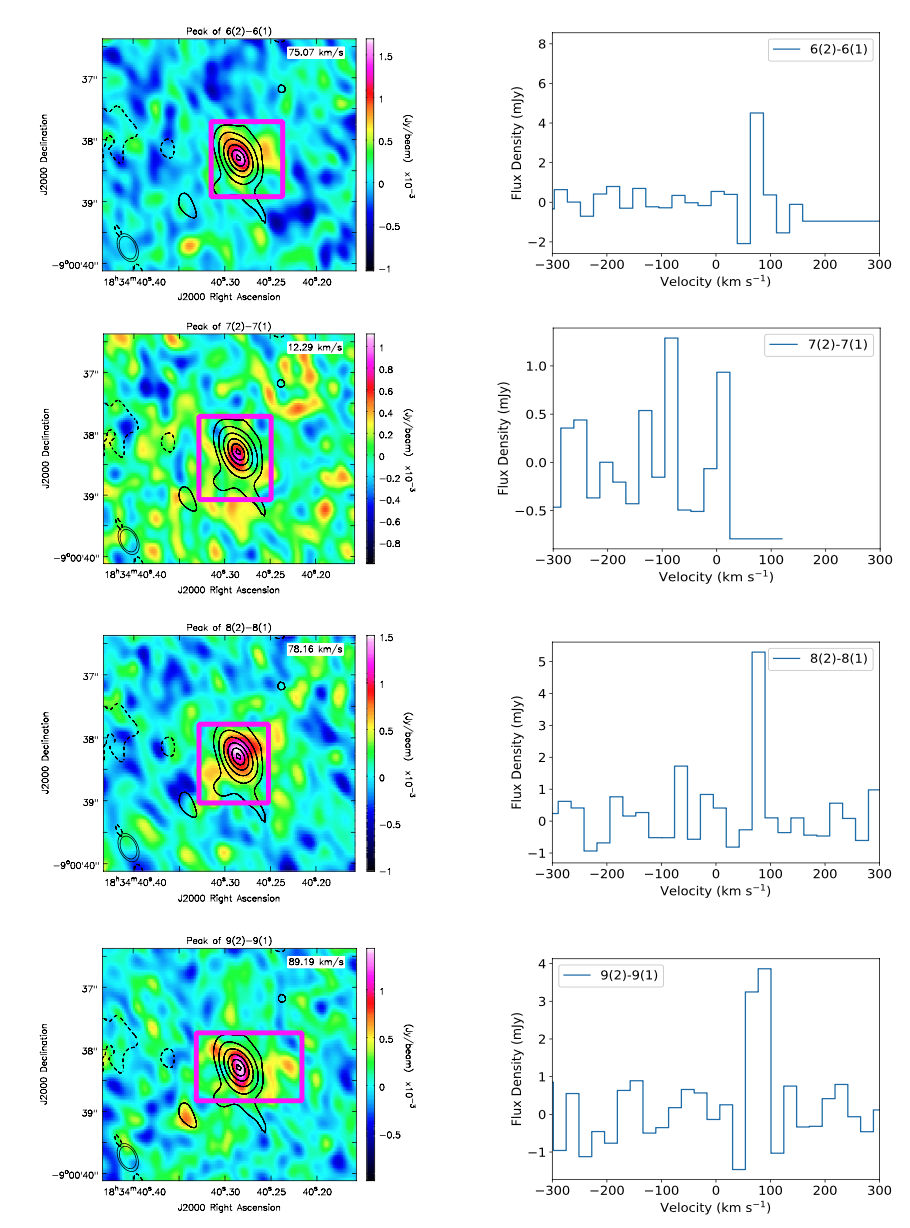}
\caption{As Figure \ref{fig_peak_CH3OH_1} but for G23.01$-$0.41A. The 1.3$\,$cm radio continuum contour levels are [$-$1.5, 3, 10, 25, 45, 65, 80] times the $9\mu$Jy$\,$b$^{-1}$ RMS noise. The SPW of the 7(2)-7(1) transition did not cover the systemic velocity of the region.}
\label{fig_peak_CH3OH_3}
\end{figure}

\begin{figure}[h]
\hspace*{1cm}
\includegraphics[scale=0.50]{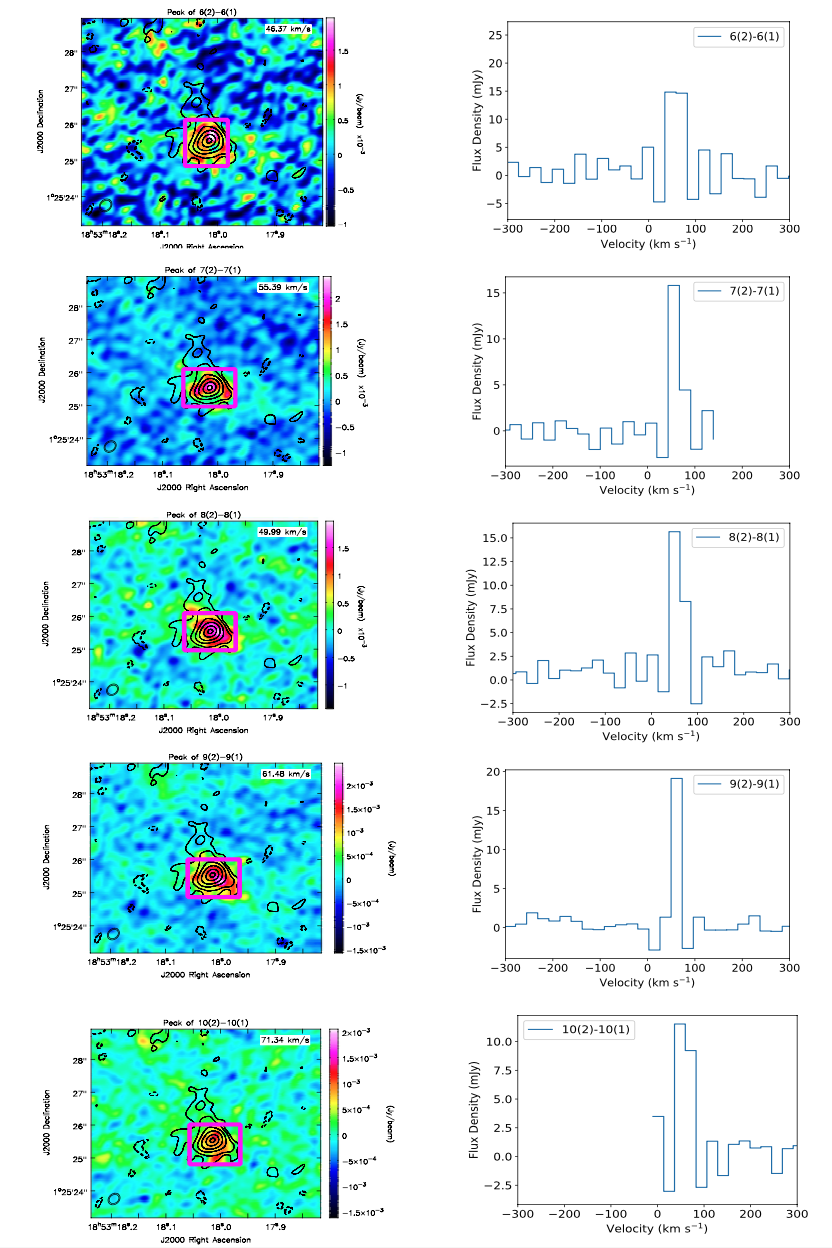}
\caption{As Figure \ref{fig_peak_CH3OH_1} but for G34.43+00.24mm1A. The 1.3$\,$cm radio continuum contour levels are  [$-$2, 3, 6, 10, 20, 40, 60] times the $10\mu$Jy$\,$b$^{-1}$ RMS noise.}
\label{fig_peak_CH3OH_3}
\end{figure}

\begin{figure}[h]
\hspace*{1cm}
\includegraphics[scale=0.52]{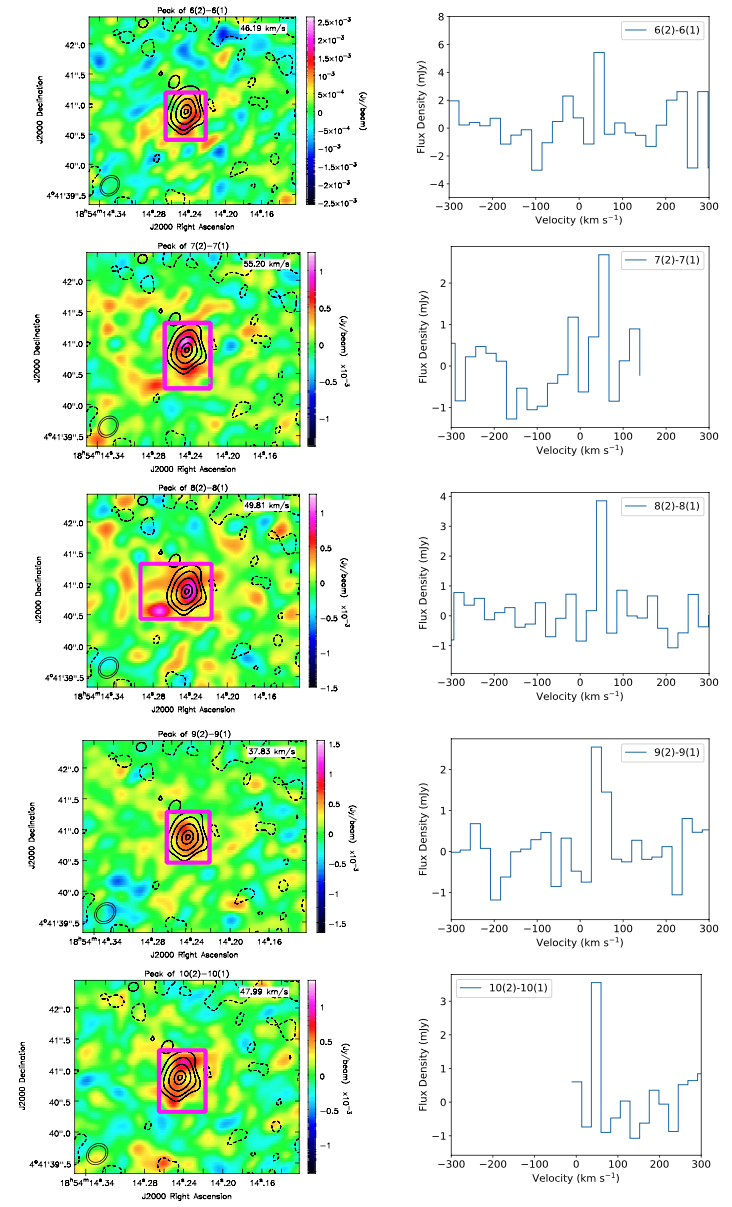}
\caption{As Figure \ref{fig_peak_CH3OH_1} but for IRAS 18517+0437 A. The 1.3$\,$cm radio continuum contour levels are [$-$1, 3, 6, 12, 20, 24] times the $8.5\,\mu$Jy$\,$b$^{-1}$ RMS noise. \cite{Rodriguez-Garza_2017ApJS..233....4R} and \cite{Gomez-Ruiz_2016ApJS..222...18G} report 44$\,$GHz CH$_3$OH masers toward this region,  but offset by more than 5$\arcsec$ in a NE-SW orientation. }
\label{fig_peak_CH3OH_4}
\end{figure}

\begin{figure}[h]
\hspace*{1cm}
\includegraphics[scale=0.75]{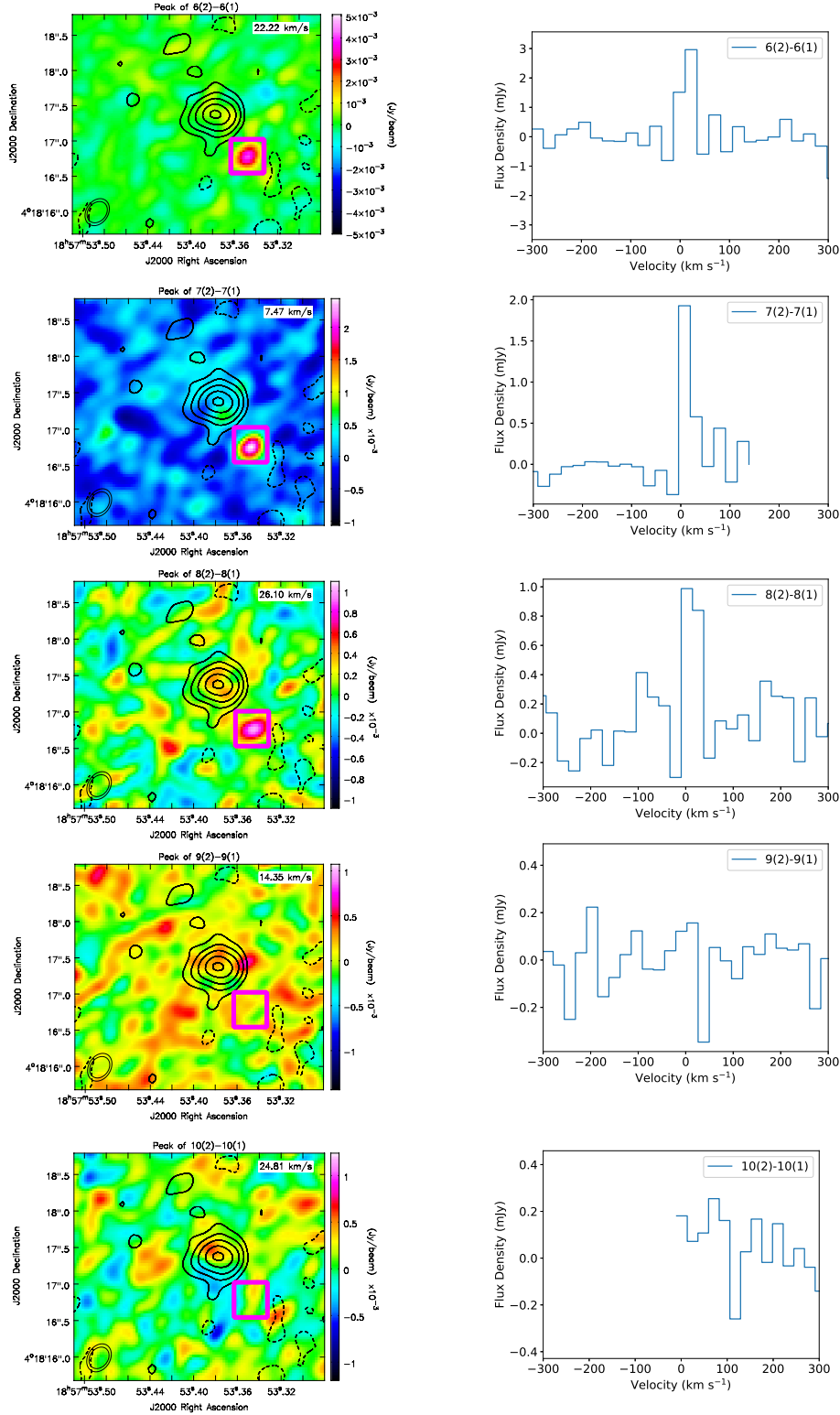}
\caption{As Figure \ref{fig_peak_CH3OH_1} but for IRAS$\,$18553+0414 A. The 1.3$\,$cm radio continuum contour levels are [$-$1.5, 3, 7, 13, 21, 28] times the $7\,\mu$Jy$\,$b$^{-1}$ RMS noise. \cite{Rodriguez-Garza_2017ApJS..233....4R} reported a 44$\,$GHz CH$_3$OH maser coincident with the detections shown in the upper three panels, strongly suggesting that these detections are also Class I masers. We report no detection of the 9(2)-9(1) and 10(2)-10(1) transitions.}
\label{fig_peak_CH3OH_5}
\end{figure}

\begin{figure}[h]
\hspace*{1cm}
\includegraphics[scale=0.75]{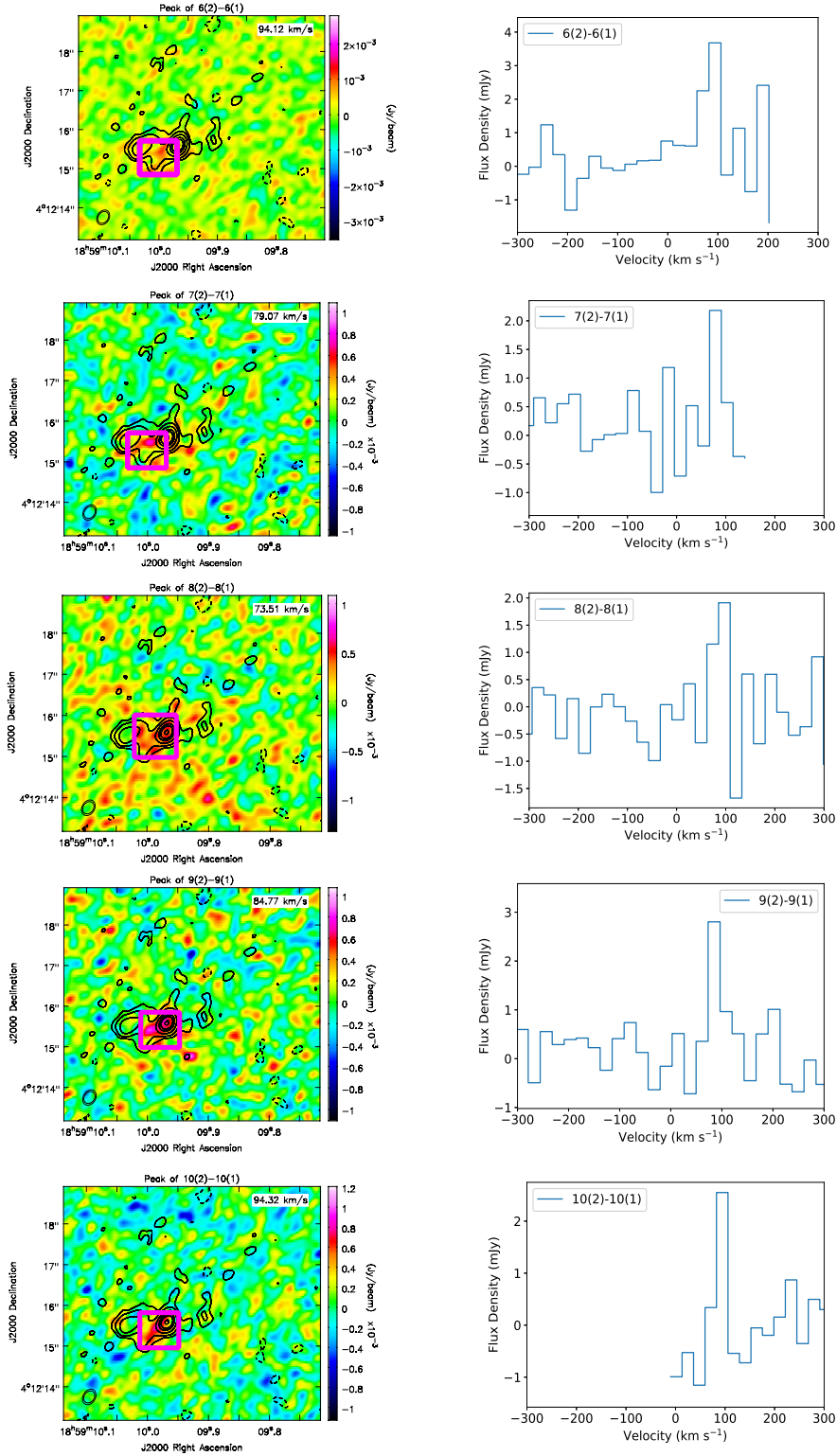}
\caption{As Figure \ref{fig_peak_CH3OH_1} but for IRAS$\,$18566+0408.  The 1.3$\,$cm radio continuum contour levels [$-$2, 3, 5, 9, 15, 20, 29] times the $7\,\mu$Jy$\,$b$^{-1}$ RMS noise. In \cite{Rodriguez-Garza_2017ApJS..233....4R}, we reported a group of three 44$\,$GHz CH$_3$OH masers $\sim 15\arcsec$ NW of the continuum source, one maser $\sim 3\arcsec$~NE, and one maser $\sim 1\arcsec$ south of the continuum source, i.e., the weak detections reported here are not coincident with 44$\,$GHz CH$_3$OH masers from our previous observations.}
\label{fig_peak_CH3OH_6}
\end{figure}

\begin{figure}[h]
\hspace*{1cm}
\includegraphics[scale=0.9]{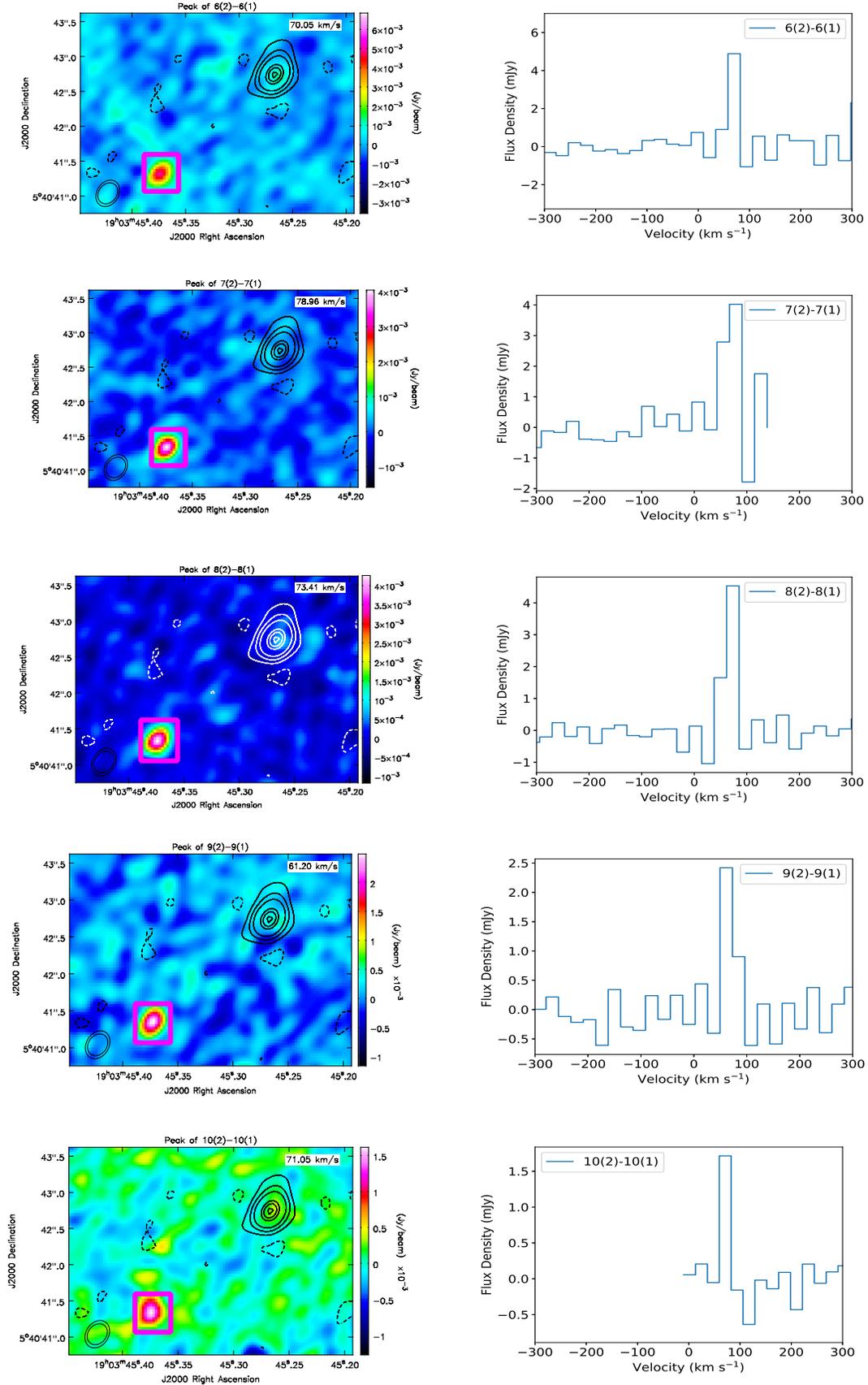}
\vspace{-1cm}
\begin{minipage}[c]{1.0\linewidth}
\vspace*{-3cm}
\caption{As Figure \ref{fig_peak_CH3OH_1} but for IRAS 19012+0536A. The 1.3$\,$cm radio continuum contour levels are [$-$2, 9, 20, 35, 75, 90, 105] times the $7.5\,\mu$Jy$\,$b$^{-1}$ RMS noise.}
\end{minipage}
\label{fig_peak_CH3OH_7}
\end{figure}

\begin{figure}[h]
  \hspace*{0.5cm}
  \vspace*{3cm}
\includegraphics[scale=0.75]{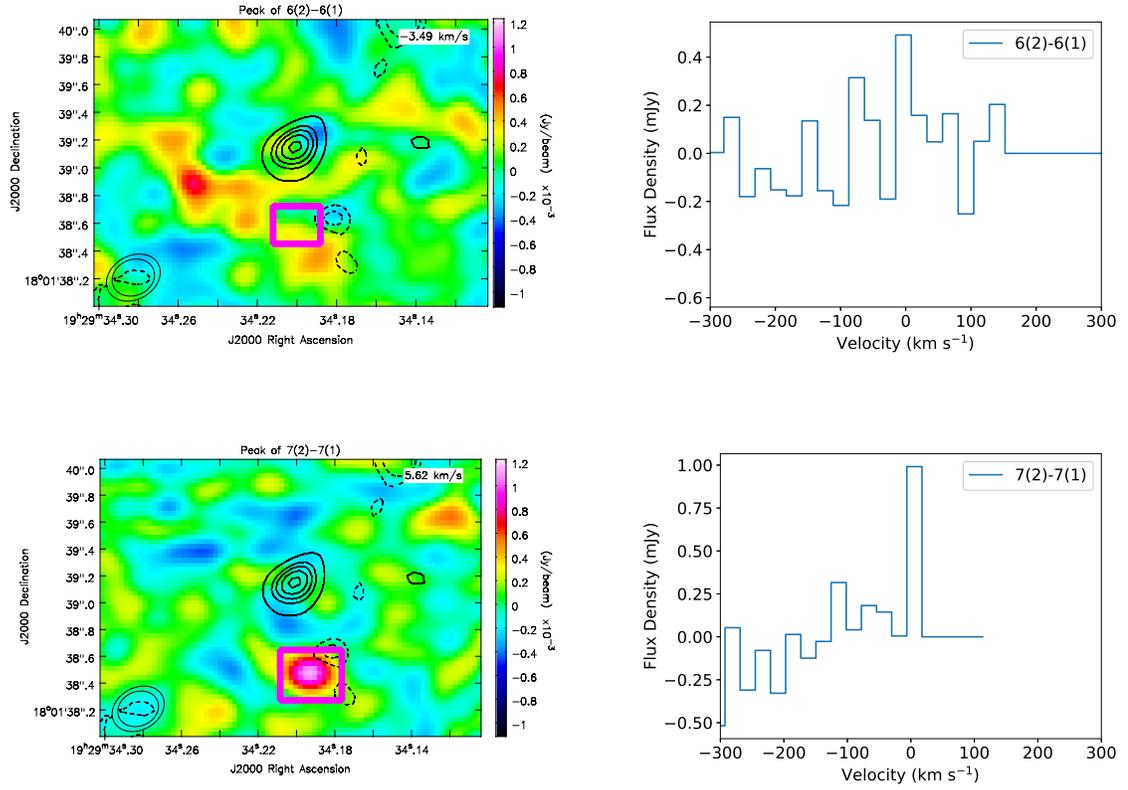}
\begin{minipage}[c]{1.0\linewidth}
\vspace*{-13cm}
\caption{As Figure \ref{fig_peak_CH3OH_1} but for G53.25$+$00.04mm4A. The 1.3$\,$cm radio continuum contour levels are [$-$2, $-$1.5, 3, 5, 6, 7, 8] times the $9\,\mu$Jy$\,$b$^{-1}$ RMS noise. We report no detection of the 6(2)-6(1) transition.}
\end{minipage}
\label{fig_peak_CH3OH_8}
\end{figure}

\begin{figure}[h]
\hspace*{1cm}
\includegraphics[scale=0.75]{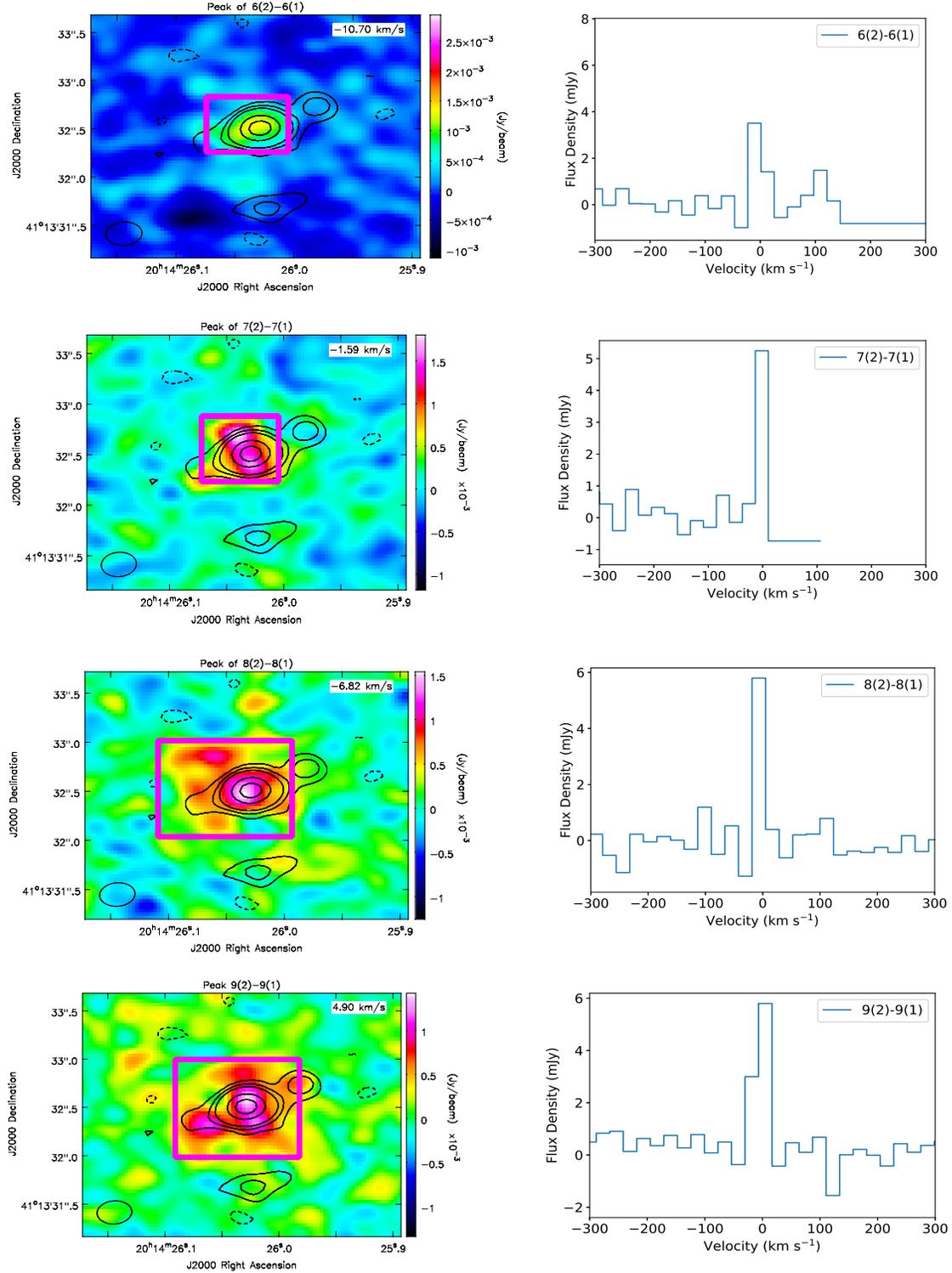}
\caption{As Figure \ref{fig_peak_CH3OH_1} but for IRAS$\,$20126+4104. The 1.3$\,$cm radio continuum contour levels are [$-$2, 3, 6, 10, 30, 50] times the $10\,\mu$Jy$\,$b$^{-1}$ RMS noise. In \cite{Gomez-Ruiz_2016ApJS..222...18G} and \cite{Rodriguez-Garza_2017ApJS..233....4R}, we report 44$\,$GHz CH$_3$OH masers offset by more than $\sim 8\arcsec$ SE and NW of the continuum source shown here. }
\label{fig_peak_CH3OH_9}
\end{figure}
\clearpage

\begin{figure}
\includegraphics[width=1.0\textwidth]{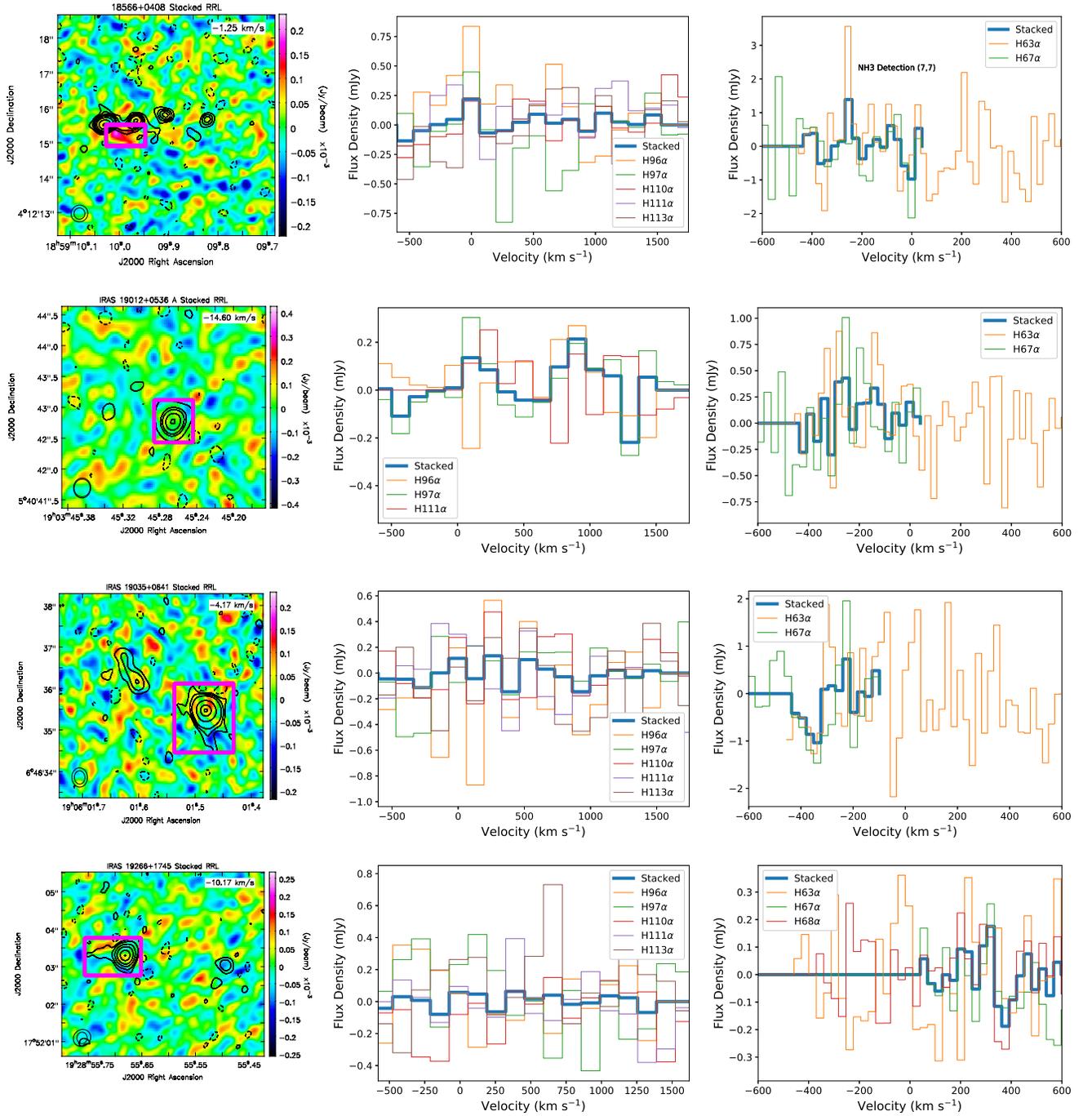}
\begin{minipage}[c]{1.0\linewidth}
\vspace*{-5cm}
\caption{{\it Left Panels}: Examples of channel maps (colors) of stacked C-band RRL cubes for four sources, superimposed to 1.3$\,$cm continuum from \cite{Rosero_2016ApJS..227...25R} (black contours). Magenta boxes are the regions used to obtain the spectra shown in the center and right panels. The synthesized beams of the channel and continuum data are shown in the lower-left corner of the images. {\it Center Panels}: Spectra of all SPWs that contain RRL frequencies observed at C-band. The thick blue curve shows the stacked spectra. {\it Right Panels}: Same as center panels but for K-band. A peak in the H63$\alpha$ spectrum of IRAS$\,$18566+0408 (upper right panel) corresponds to a NH$_3$ transition. The stacked spectra were obtained after the spectra of the individual transitions were regrid and aligned in velocity; we show the K-band RRL spectra of individual RRL transitions before regrid.}
\end{minipage}
\label{fig_RRL}
\end{figure}

\begin{figure}[h]
\centering
\begin{minipage}[t]{0.45\textwidth}
  \scalebox{1.5}{\includegraphics[width=0.72\textwidth]{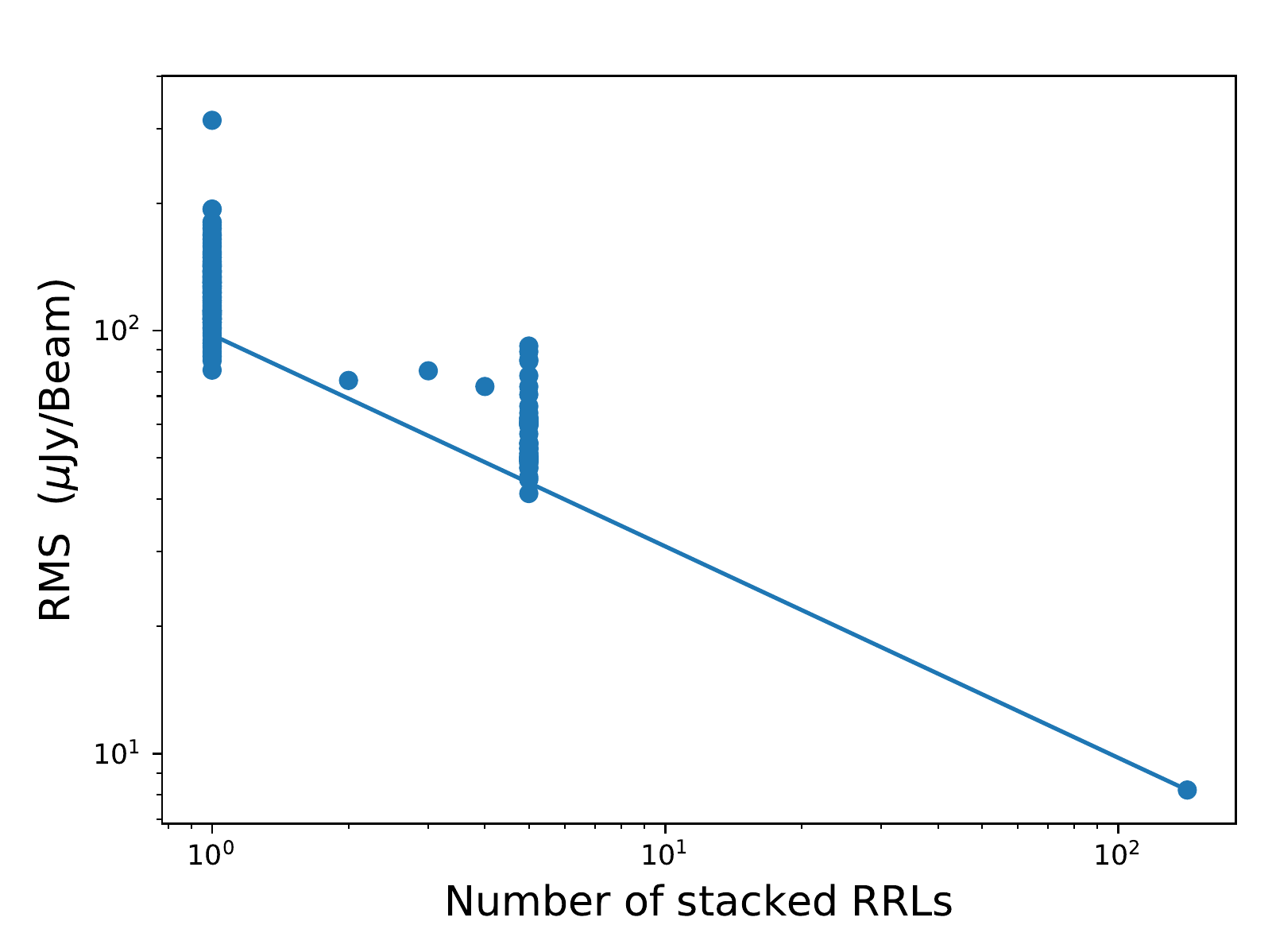}}
 \end{minipage}
\hfill
\begin{minipage}[t]{0.45\textwidth}
  \hspace*{-1cm}
  \scalebox{1.5}{\includegraphics[width=0.72\textwidth]{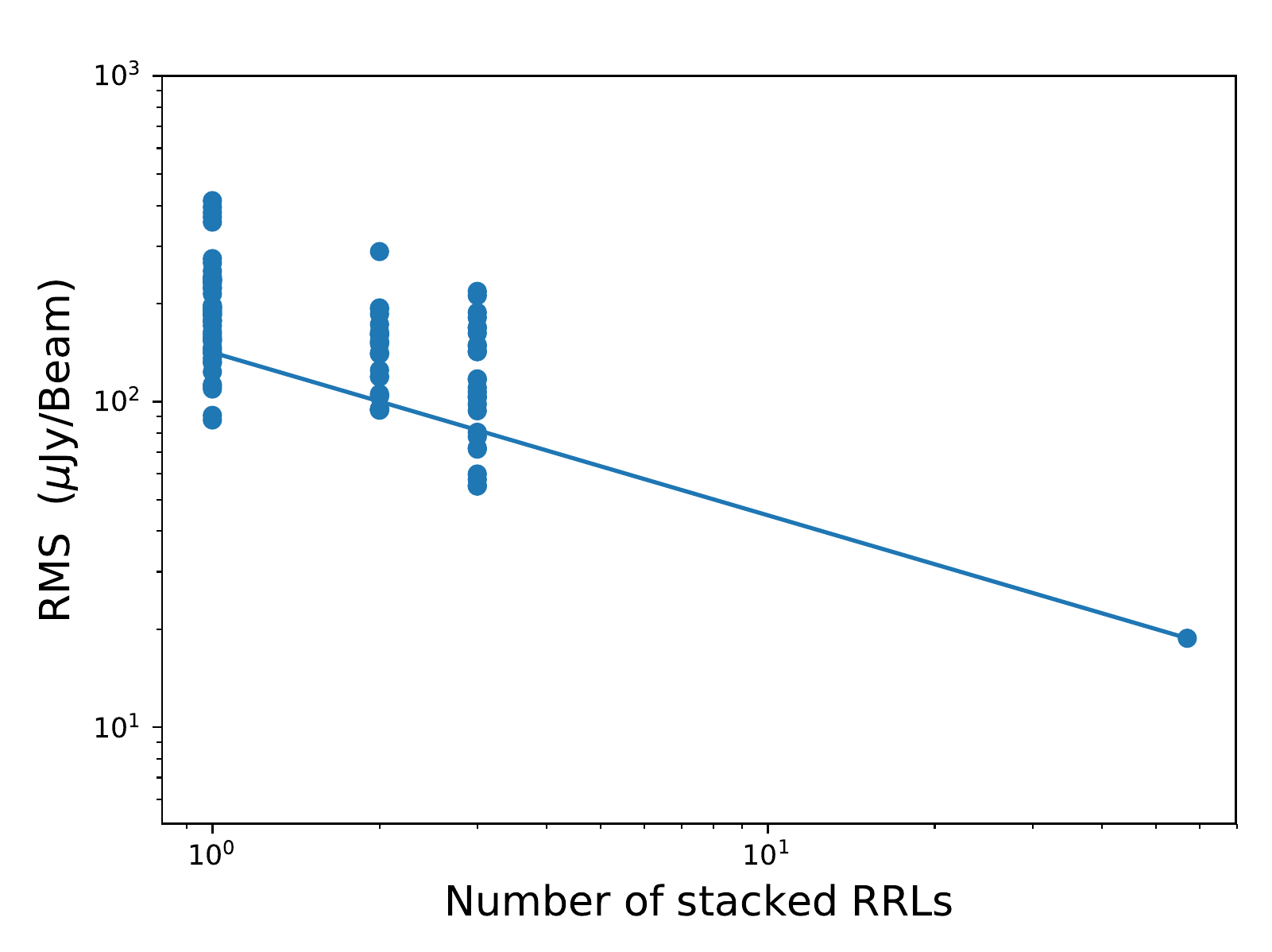}}
 \end{minipage}
\caption{ {\it Left Panel:} RMS values as function of number of stacked RRLs at C-band (6$\,$cm). The data corresponding to $10^0$ are the RMS values of the individual RRL spectral cubes, the data between 2 and 5 stacked RRLs correspond to the RMS values when all RRL spectra of the same source are averaged. The data point to the right of the graph corresponds to the RMS of the cube for which all sources and transitions were stacked. The blue line is the expected sensitivity function from the radiometer equation ($RMS\, \propto N^{-1/2}$) extrapolated from the rightmost data point. {\it Right Panel:} Same as left panel but for K-band (1.3$\,$cm) data.\label{fig_ResultRC} }
\end{figure}

\end{document}